\documentclass[showpacs,aps,prb,twocolumn,superscriptaddress]{revtex4-1}

\usepackage[english]{babel}
\usepackage[pdftex]{graphicx}
\usepackage{hyperref}
\usepackage{epsfig}
\usepackage{color}
\usepackage{psfrag}
\usepackage{float}
\usepackage{bbold}
\usepackage{tikz}
\usepackage{amsmath}
\usepackage{amssymb}

\definecolor{oldgold}{rgb}{0.81, 0.71, 0.23}
\definecolor{richelectricblue}{rgb}{0.03, 0.57, 0.82}

\newcommand{\ket}[1] {\left|#1\right\rangle}
\newcommand{\bra}[1] {\langle #1 |}

\begin{document}

\title{Charge-Transfer Chemical Reactions in Nanofluidic  Fabry-P\'{e}rot Cavities}

\author{L. Mauro}
\affiliation{Univ. Bordeaux, CNRS, LOMA, UMR 5798, F-33400 Talence, France}
\author{K. Caicedo}
\affiliation{Univ. Bordeaux, CNRS, LP2N, UMR 5298, F-33400 Talence, France}
\author{G. Jonusauskas}
\affiliation{Univ. Bordeaux, CNRS, LOMA, UMR 5798, F-33400 Talence, France}
\author{R. Avriller}
\email{remi.avriller@u-bordeaux.fr}
\affiliation{Univ. Bordeaux, CNRS, LOMA, UMR 5798, F-33400 Talence, France}

\date{\today}

\begin{abstract}
We investigate the chemical reactivity of molecular populations confined inside a nanofluidic Fabry-P\'{e}rot cavity.
Due to strong light-matter interactions developing between a resonant electromagnetic cavity-mode and the electric dipole moment of the confined molecules, a polariton is formed.
The former gets dressed by environmental vibrational and rotational degrees of freedom of the solvent.
We call the resulting polariton dressed by its cloud of environmental excitation a ``\textit{reacton}", since it further undergoes chemical reactions.
We characterize how the \textit{reacton} formation modifies the kinetics of a photoisomerization chemical reaction involving an elementary charge-transfer process. 
We show that the reaction driving-force and reorganization energy are both modulated optically by the reactant concentration, the vacuum Rabi splitting and the detuning between the Fabry-P\'{e}rot cavity frequency and targeted electronic transition.
Finally, we compute the ultrafast picosecond dynamics of the whole photochemical reaction.
We predict that despite optical cavity losses and solvent-mediated non-radiative relaxation, measurable signatures of the \textit{reacton} formation can be found in state-of-the-art pump-probe experiments.
\end{abstract}

\maketitle

\section{Introduction}
\label{Introduction}
Electron-transfer (ET) chemical reactions in solution constitute a paradigmatic class of chemical reactions \cite{pauling1988general}.
In the simplest case, an electron of charge $-e$ ($e$ is the elementary charge) is transferred from an anion $(A^-)$ to a cation $(C^+)$, following the ET reaction $A^- + C^+ \longrightarrow A + C$.
In the more general class of charge-transfer (CT) chemical reactions, a modification of the local-charge density of states occurs between different chemical groups of the reacting molecules, thus resulting in a partially-transferred (shifted) charge $\delta_{\rm{e}}$ during the CT process.
Such is the case for intramolecular CT reactions in $D-A$ molecules where an electron donor (D) group is connected to an electron acceptor (A) group through a molecular bridge ($-$), thus resulting in the following intramolecular CT mechanism $D-A \longrightarrow D^{+\delta_{\rm{e}}}-A^{-\delta_{\rm{e}}}$.
The theoretical description of ET and CT reactions has a long history \cite{pauling1988general}, which was fully developed after the works of Marcus \cite{marcus_theory_1956,marcus_electron_1993,siders_quantum_1981,marcus_nonadiabatic_1984}, and Kestner et al. \cite{kestner_thermal_1974}, with later successful applications for biological molecules \cite{hopfield_electron_1974}. 
At the heart of Marcus theory is the necessity to take into account explicitly the solvent in the modelling of ET reaction rates.
Out-of-equilibrium fluctuations in the solvent nuclear coordinates are indeed necessary to reach the crossing point of the reactant (R) and product (P) potential energy surfaces (PES), at which the electron-transfer occurs. 
The reaction rate $k_{\rm{ET}}$ is given by the simple result of Marcus \cite{marcus_theory_1956}
$k_{\rm{ET}}=k_e \exp{ \left( -\Delta_r G^*/k_B T \right)}$, with $k_e$ a reaction-dependent global rate, $T$ the temperature, $k_B$ the Boltzmann constant and $\Delta_r G^* = \left( \Delta_r G^0 + \lambda_{\rm{S}} \right)^2/4\lambda_{\rm{S}}$ an effective activation energy depending on the solvent reorganization energy $\lambda_{\rm{S}}$ and variation of the thermodynamic Gibbs potential $\Delta_r G^0$ (reaction driving-force).
Although this expression takes a form similar to transition-state theory \cite{eyring_activated_1935,evans_applications_1935,wigner_transition_1938,kramers_brownian_1940}, 
it is important to notice that its derivation using quantum mechanical first principles does not involve the concept of activated-complex or transition-state \cite{kestner_thermal_1974,siders_quantum_1981}.
%

%
The experimental investigation of a wide variety of ground-state chemical reactions, photoreactions, ET and CT reactions came recently to a strong revival, due to the ability of confining molecular ensembles inside micro or nano-optical and plasmonic cavities \cite{ebbesen_hybrid_2016,ribeiro2018polariton}, resulting in an alteration of their chemical reactivity.
%
For instance, it was shown that electromagnetic microcavities, can be tailored such that a single cavity-mode of frequency $\omega_c \approx 2.2 \mbox{ eV}/\hbar$ (with $\hbar$ the Planck constant) can be tuned to resonance with the electronic transition between the ground and excited state of the cavity-confined molecules \cite{schwartz_reversible_2011,hutchison_modifying_2012}.  
The conjunction of a low cavity volume $V$, large number of confined molecules $N$ and strong molecular electric dipole $\mu$,  
results in sizable vacuum quantum fluctuations of the 
cavity electrical field $E_0=\sqrt{\hbar\omega_c/2\varepsilon_0 V}$, with $\varepsilon_0$ the electromagnetic vacuum permittivity. 
The resulting light-matter coupling strength between the molecular dipoles and the cavity mode, as quantified by the collective vacuum Rabi splitting frequency $\tilde{\Omega}_R=\mu E_0 \sqrt{N}/\hbar$ \cite{haroche_cavity_nodate,dicke_coherence_1954,tavis_exact_1968}, can be as high as $\tilde{\Omega}_R \approx 0.7 \mbox{ eV}/\hbar$ \cite{schwartz_reversible_2011}, thus exceeding the cavity optical losses $\kappa \approx 0.2 \mbox{ eV}/\hbar$.
In this regime of electronic light-matter strong coupling, a collective hybrid excitation is formed between the resonant cavity-mode and the embedded molecules called \textit{polariton}.
As was reported and investigated in depth for optical spectra in semiconducting microcavities \cite{weisbuch_observation_1992,houdre_early_2005}, polariton excitations are
characterized by the vacuum Rabi-splitting of cavity optical absorption spectra \cite{schwartz_reversible_2011}.
It is remarkable that the strong-coupling regime (with $\tilde{\Omega}_R \approx 110 \mbox{ meV}/\hbar$ and $\kappa \approx 60 \mbox{ meV}/\hbar$) was also recently achieved in liquid phase, in which optically active molecules are confined inside a nanofluidic Fabry-P\'{e}rot cavity \cite{bahsoun_electronic_2018}.
%

The formation of cavity polaritons has deep consequences on the chemical 
reactivity of the embedded molecules.
It was shown experimentally that the potential energy landscape of a photoisomerization chemical reaction is strongly altered under resonant conditions between an electromagnetic cavity mode and the electronic transition between ground and excited states of the reaction product \cite{schwartz_reversible_2011}, thus resulting in a significant slowing-down of the reaction kinetics \cite{hutchison_modifying_2012}. 
Theoretical investigations have described and computed the polariton potential energy surface (PPES) of such class of reactions, taking into account the role of nuclear degrees of freedom in describing the light-matter interaction mechanism \cite{galego_cavity-induced_2015}, the former inducing a vibrational dressing of the polaritons \cite{cwik_excitonic_2016}.
The resulting alteration of the PPES was shown to be responsible for the cavity-induced slowing-down of photochemical reactions \cite{galego_suppressing_2016,galego_many-molecule_2017}.
On the other hand, ET chemical reactions for molecular populations in cavity were also predicted to be accelerated by orders of magnitude, as a result of both the modulation of the PPES and the influence of the collective decoupling between nuclear motion and electronic degrees of freedom in the light-matter strong-coupling regime \cite{herrera_cavity-controlled_2016}.  
Recent theoretical works on ET reactions in confined electromagnetic environment reported a cavity-induced significant enhancement of the ET reaction rate \cite{semenov2019electron}, and
emphasize the role of counter-rotating terms \cite{mandal2020polariton} (beyond rotating-wave approximation) and self-dipole energy terms \cite{semenov2019electron,mandal2020polariton} in writing the interaction Hamiltonian: both are necessary for preserving gauge invariance \cite{craig1998molecular,di2019resolution} and computing accurately the PPES upon entering the ultra-strong coupling regime of cavity quantum electrodynamics ($\tilde{\Omega}_R \geq \omega_c$).
%

In this paper, we revise the theoretical description of the kinetics of CT chemical reactions in nanofluidic Fabry-P\'{e}rot cavities. 
We investigate on the same footing the collective coupling between molecular populations and a single electromagnetic cavity-mode, taking into account dissipation and dephasing mechanisms induced by the solvent and cavity losses. 
%

The organization of the paper is the following. 
In Sec.\ref{Theoretical_Modelling}, we introduce our theoretical microscopic model of solvated molecules interacting with a single electromagnetic cavity-mode. 
We develop an analytical scheme based on the Born-Oppenheimer approximation that enables to compute analytically the PPES in the regime for which the collective vacuum Rabi splitting $\tilde{\Omega}_R$ is larger than the intramolecular vibrational reorganization energy $\lambda_{\rm{v}}$.
In this regime, we obtain approximate many-body wave functions for the polaritons and dark states of the molecules-cavity ensemble, in presence of coupling to the reaction coordinate and solvent bath.
In some limits, we recover the results of Refs.\cite{herrera_cavity-controlled_2016,wu2016polarons,zeb2017exact}, that are based on the use of the variational polaron ansatz \cite{toyozawa1954theory,silbey_electronic_1976,bera2014generalized}.
We interpret physically this result by introducing the concept of \textit{reacton}, which is the collective excitation of the reactant molecules interacting strongly with the cavity-mode and dressed by its interaction with the solvent. 
In Sec.\ref{Charge_transfer_reaction_rate}, we derive a generalization of Marcus theory \cite{marcus_electron_1993} adapted to the \textit{reacton}'s formation inside the electromagnetic cavity. 
We improve the already existing theory of Ref.\cite{herrera_cavity-controlled_2016} by adapting a theoretical framework derived by Kestner et al. \cite{kestner_thermal_1974} for describing ET reactions in solution.
This enables to incorporate explicitly the solvent into the reaction mechanism by using the separation of time-scales between fast intra-molecular vibrational modes along the reaction coordinate and slow vibrational modes of the solvent bath. 
Compared to more recent Refs.\cite{semenov2019electron,mandal2020polariton}, we improve several points of the theory by including explicitly both the collective coupling of N molecules to the cavity-mode (and not of a single molecule) and the presence of dissipation by the environment. 
We then compute the modification of the CT reaction rate due the formation of the \textit{reacton} inside the cavity, for a specific model of photoreaction involving a charge-transfer process in the electronic excited-state.
We show that the \textit{reacton} opens new channels for the charge-transfer mechanism. 
Depending on the range of parameters, the reaction kinetics can either be slower or faster inside cavity compared to outside cavity. 
In Sec.\ref{Dissipation}, we derive the dissipation and dephasing rates induced by the cavity optical losses, non-radiative relaxation induced by molecular vibrations, and dephasing of the \textit{reacton} by the solvent bath. 
For this purpose, we extend the approach derived from quantum optics in Ref.\cite{canaguier-durand_non-markovian_2015} using the 
the dressed-atom approach \cite{cohen1998atom}, to our case of many-body \textit{reacton} basis.
In Sec.\ref{Ultrafast_reaction_kinetics}, we solve numerically the 
whole ultrafast picosecond kinetics of the photoreaction.
We develop a rate-equation approach that we solve numerically, obtaining the 
time-dependent evolution of reactants and products concentration inside the cavity, after a 
single-photon has been absorbed to initiate the reaction.
Despite strong cavity losses and dissipation induced by the solvent, we predict fingerprints of the \textit{reacton} formation that should be visible on picosecond time-scales. 
Finally, we develop in Sec.\ref{Perspectives} some open perspectives in this field that are of interest for the design and engineering of a new generation of open chemical reactors, the kinetics of which is modulated by vacuum quantum fluctuations of the cavity electromagnetic field. 
%

\section{Theoretical modelling}
\label{Theoretical_Modelling}

\subsection{Microscopic Hamiltonian}
\label{Microscopic_Hamiltonian}
%
%
\begin{figure}[!h]
    \centering
    \includegraphics[width=1.0\linewidth]{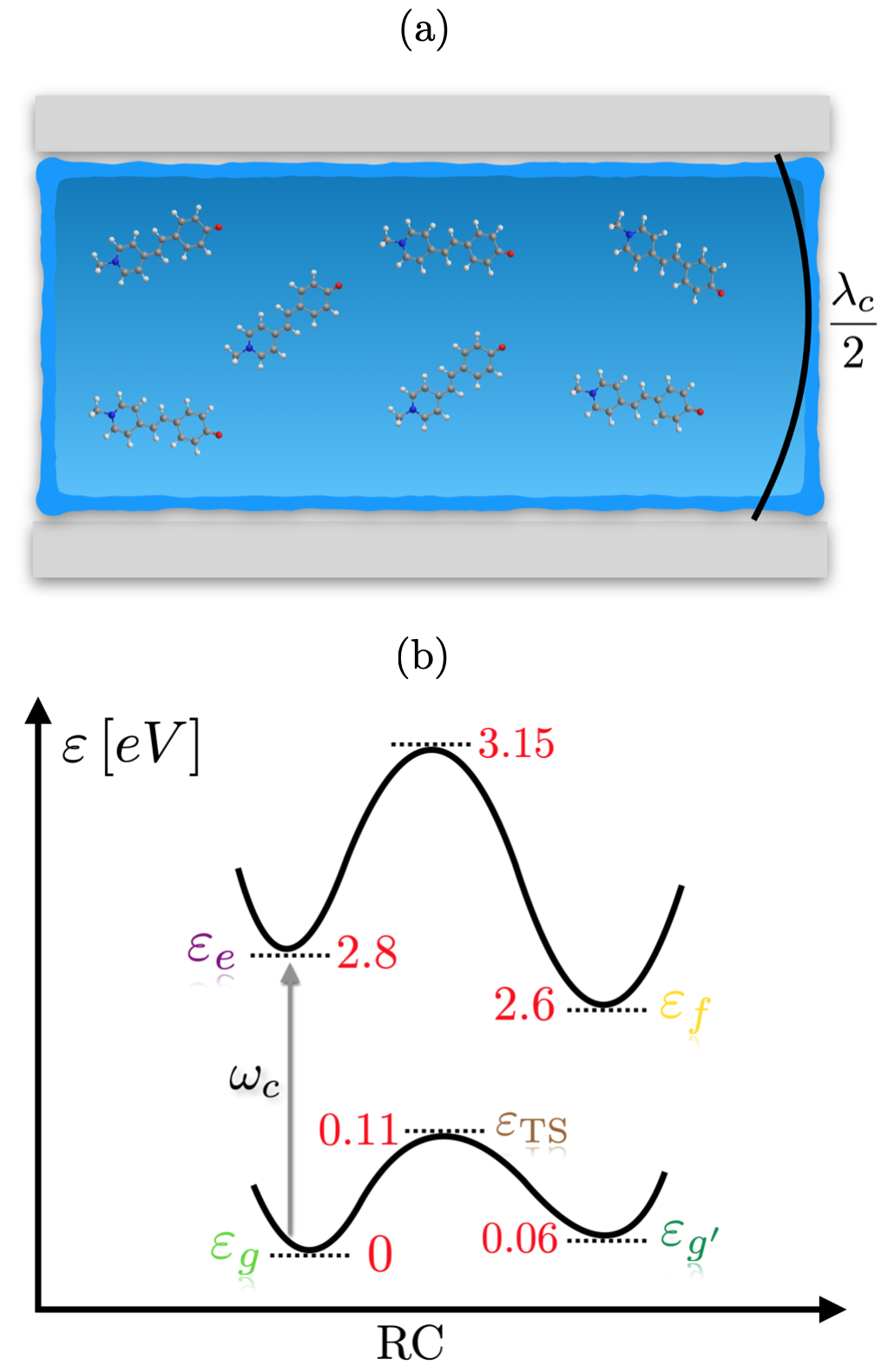}
    	\caption{
    (a) Pictorial representation of molecules of (E)-4-(2-(1-methylpyridin-1-ium-4-yl)vinyl)phenolate, in solution inside a nanofluidic Fabry-P\'{e}rot cavity.
        The nomenclature describes this photoactive molecule in its aromatic form.
        %
        %
        %
        $\lambda_c/2 = \pi c/ n \omega_c$ is the wavelength of the cavity fundamental electromagnetic mode, with $c$ the speed of light and $n$ the refractive index of the medium.
        (b) Sketch of the PES for such molecules as a function of the RC. 
        The electronic ground-state minima $g$ and $g'$ and excited-state minima $e$ and $f$ for the molecule are presented as well as their typical energies $\varepsilon_g$, $\varepsilon_{g'}$, $\varepsilon_e$ and $\varepsilon_f$ in eV.
        The grey arrow stands for the cavity-mode of frequency $\omega_c$ that is resonant with the $g-e$ electric dipole transition.
       }
    \label{fig:Fig1}
\end{figure}

We investigate the chemical reactivity of a solution of molecules inside a Fabry-P\'{e}rot nanofluidic cavity. 
For this purpose, bi-phenyl molecules have been studied extensively \cite{maus_excited_2002,herrera_cavity-controlled_2016}, since they have interesting photochemical properties due to a rotational degree of freedom around a C-C bond connecting the phenyl groups, as well as a possibility of being functionalized by various chemical groups with electron donating or accepting character. 
Other donor-acceptor molecules with an internal high-frequency vibrational mode are also good candidates for investigating CT reaction rates in solution.
In our paper, we consider typical organic molecules with interesting photoactive properties, embedded inside the cavity.
Such is the case for the molecule represented in Fig.\ref{fig:Fig1}a, and written (E)-4-(2-(1-methylpyridin-1-ium-4-yl)vinyl)phenolate; this nomenclature describes the structure of the molecule in its aromatic form. 
We show in Fig.\ref{fig:Fig1}b a sketch of the PES for such a molecule described within Born-Oppenheimer approximation \cite{tully_perspective_2000}, as a function of the reaction coordinate (RC).
The RC corresponds to an intra-molecular vibration or a rotation mode of the molecule.
The electronic structure of this molecule is described by an electronic ground-state with two relative minima labelled $g$ and $g'$, and an electronic excited-state with two minima $e$ and $f$.
Upon photoexcitation from $g$ to $e$, the molecule can reach the more stable excited-state $f$, 
by changing its conformation and undergoing an elementary CT process.
For simplicity, we approximate the complex electronic structure of the molecule by displaced parabolic PES \cite{herrera_cavity-controlled_2016}, in the spirit of the parabolic approximation in Marcus theory \cite{marcus_electron_1993}.
We consider the system made of N molecules in solution coupled to a single electromagnetic cavity-mode (see Fig.\ref{fig:Fig1}a). 
We write the microscopic Hamiltonian $\mathcal{H}$ describing this system
\begin{eqnarray}
\mathcal{H} = H_{\rm{CaM}} + V_{\rm{M-Ca}} + V_{\rm{CT}}
\label{H_CaM1}\,,
\end{eqnarray}
as the sum of the Hamiltonian $H_{\rm{CaM}}$ describing the free electromagnetic cavity-mode (Ca) and quadratic PES of the solvated molecules (M), plus the Hamiltonian $V_{\rm{M-Ca}}$ standing for electromagnetic interactions between the molecules and the cavity-mode. 
We denote $V_{\rm{CT}}$ the Hamiltonian describing weak-coupling between 
electronic excited-states $e$ and $f$ of the molecule, at the origin of charge-transfer.
Each of those Hamiltonian is given by
\begin{eqnarray}
&&H_{\rm{CaM}} = \sum_{i=1}^{N} \sum_{r=g,g',e,f} \varepsilon_{ri}
\ket{r_i} \bra{r_i} + \hbar\omega_c\left( a^\dagger a + \frac{1}{2} \right)
\,,\label{H_CaM2} \\
&&\varepsilon_{ri} = \varepsilon_{r} + \frac{\omega_\mathrm{v}^2}{2}\left(
Q_{\mathrm{v},i} - \overline{Q}_{\mathrm{v},r}\right)^2
+
\sum_{k}\frac{\omega_k^2}{2}\left(
Q_{\mathrm{S},ik} - \overline{Q}_{\mathrm{S},rk}\right)^2
\,, \nonumber \\
\label{H_CaM3}\\
&&V_{\rm{M-Ca}} = \frac{\hbar\Omega_R}{2}\sum_{i=1}^{N} \left( 
\ket{e_i} \bra{g_i} a 
+
a^\dagger \ket{g_i} \bra{e_i}
\right)
\,,\label{H_CaM4} \\
&&V_{\rm{CT}} = \sum_{i=1}^{N} \left( 
V_{ef} \ket{e_i} \bra{f_i} 
+
V^*_{ef} \ket{f_i} \bra{e_i} 
\right)
\,,\label{H_CaM5}
\end{eqnarray}
with $\varepsilon_{ri}$ the PES corresponding to $\ket{r_i}$ the electronic state $r=g,g',e,f$ belonging to the molecule number $i=1,\cdots,N$.
The PES in Eq.\ref{H_CaM3} is the sum of an electronic part $\varepsilon_{r}$ (bottom of the parabola in Fig.\ref{fig:Fig1}b), plus a quadratic dependence along the nuclear coordinate $Q_{\mathrm{v},i}$ corresponding to the intra-molecular vibration mode of molecule $i$, plus molecular vibrations $Q_{\mathrm{S},ik}$ of the bath of solvent molecules labelled with a quasi-continuum index $k$.
We suppose that each molecule has the same intra-molecular vibration frequency $\omega_{\mathrm{v}}$ and bath mode frequency $\omega_k$ along the RC, independently of its electronic state $r$ (same curvature around each minimum of the bare PES in Fig.\ref{fig:Fig1}b).
We label $\overline{Q}_{\mathrm{v},r}$ and $\overline{Q}_{\mathrm{S},rk}$ the displaced nuclear equilibrium positions associated respectively to the intra-molecular and solvent modes, both depending on the electronic state $r$.
The free electromagnetic mode of the cavity is described in Eq.\ref{H_CaM2} by $a$ ($a^\dagger$) the annihilation (creation) operator of a photon excitation inside the cavity of frequency $\omega_c$.
The light-matter interaction Hamiltonian in Eq.\ref{H_CaM4} is an electric-dipole coupling term, written within rotating-wave approximation (RWA) \cite{cohen1998atom,dicke_coherence_1954,tavis_exact_1968}.
It couples the electronic ground-state $g$ to the excited-state $e$ of each molecule $i$ through the same cavity-mode, with a coupling strength given by the bare vacuum Rabi frequency $\Omega_R\equiv \mu E_0/\hbar$.
We suppose for simplicity that there is no direct dipole coupling between the $g'$ and $f$ states, either because the corresponding dipole matrix elements are weak, or the cavity frequency is detuned from the corresponding electronic transition. 
We note that counter-rotating and self-dipole energy terms have been neglected in Eq.\ref{H_CaM4}.
Those terms are derived in Ref.\cite{craig1998molecular} and their effects have been investigated in depth in recent Refs.\cite{semenov2019electron,mandal2020polariton}.
They both give rise to energy shifts of the PES of relative order $\tilde{\Omega}_R/\omega_c$ compared to the standard RWA. 
Those terms are thus weak but sizable in the strong (but not ultra-strong) coupling regime $(\hbar\kappa < \hbar\tilde{\Omega}_R < \hbar\omega_c)$. 
As a first approximation, we neglect them in the Hamiltonian, in order to be able to derive tractable analytical approximations for computing the polaritonic PES and reaction rates. 
For typical values of the collective Rabi frequency $\hbar\tilde{\Omega}_R\approx 0.2-0.7\mbox{ eV}$ and cavity frequency $\hbar\omega_c\approx 2.8 \mbox{ eV}$ in a nanofluidic cavity, the corresponding corrections are of order $7-25\%$.
Finally, the matrix element $V_{ef}$ in Eq.\ref{H_CaM5} is at the origin of the intramolecular CT process between any $e$ and $f$ state of one molecule. 
The Hamiltonian $V_{\rm{CT}}$ is supposed to be a weak perturbation to the Hamiltonian $\mathcal{H}_0= H_{\rm{CaM}} + V_{\rm{M-Ca}}$ containing the molecular population coupled to the cavity-mode, but uncoupled to the excited-states $f$ and $g'$. 
This approach holds in the incoherent regime of electron-transfer for which $|V_{ef}| \ll k_B T$.
In the following, we denote $\Delta_{gr}= \varepsilon_r - \varepsilon_g$, the 
difference of electronic energies between the molecular ground-state $g$ and the excited-state $r$. 
The detuning between the cavity-mode frequency and the targeted electronic dipole transition $g-e$ is written as $\delta = \omega_c - \Delta_{ge}/\hbar$.%

\subsection{Polaritonic Potential Energy Surfaces (PPES)}
\label{PPES}
%
%
\begin{figure}[!h]
    \centering
    \includegraphics[width=1.0\linewidth]{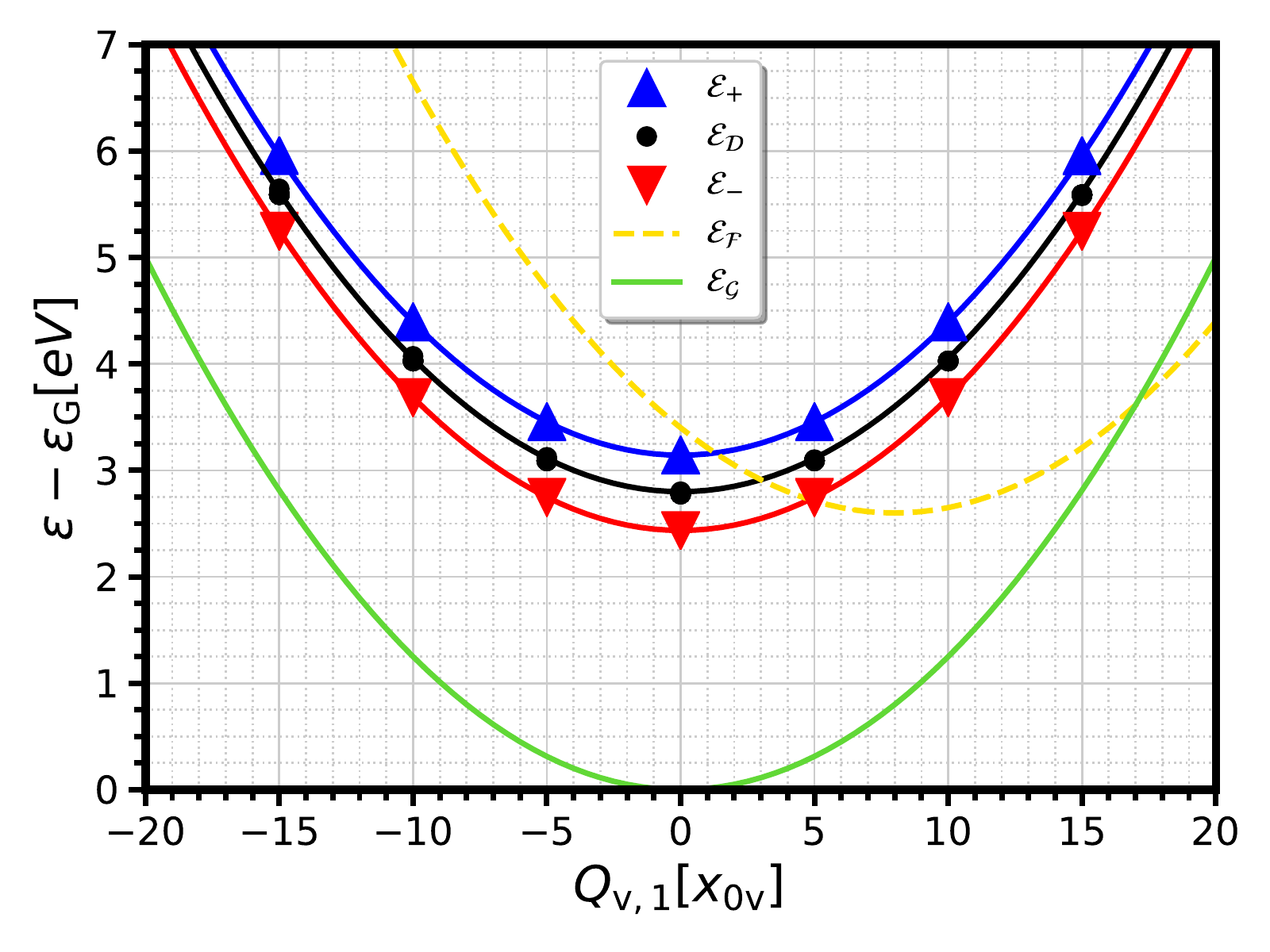}
    	\caption{
    	PPES for the lower polariton $\mathcal{E}_{-}$ (red triangle down), upper polariton $\mathcal{E}_{+}$ (blue triangle up) and dark states $\mathcal{E}_{D}$ (black dots).
    	Dotted curves are computed from numerical diagonalization of $\mathcal{H}_0$ (within RWA and absence of coupling to the solvent). 
    	The corresponding plain curves are obtained from analytical formula in Eq.\ref{PPES1} and 
    	Eq.\ref{DS2}.
    	The plain green and dashed yellow curves are the PES for the ground-state $\mathcal{G}$ and excited-state $\mathcal{F}$ respectively. 
    	Parameters are: $N = 50$, $Q_{\mathrm{v},i}$ fixed for all $i=2,\cdots, N$ with a value equals to $10 x_{0v}$ while $Q_{\mathrm{v},1}$ is varied, $\varepsilon_{g}= 0 \mbox{ eV}$, $\varepsilon_{e}= 2.8 \mbox{ eV}$, $\varepsilon_{f}= 2.6 \mbox{ eV}$, $\hbar\omega_{c}=2.8 \mbox{ eV}$, $\hbar\omega_{\mathrm{v}}=50 \mbox{ meV}$, $\hbar\Omega_R = 0.1 \mbox{ eV}$ ($\hbar\tilde{\Omega}_{R} = 0.7 \mbox{ eV}$), $\hbar\delta= 0 \mbox{ eV}$, $\lambda_{\mathrm{v},e}=0.1 \mbox{ meV}$.
    	}
    \label{fig:Fig2}
\end{figure}
In this section, we compute the polariton PES (PPES), assuming a vanishing Hamiltonian $V_{\rm{CT}}$ in Eq.\ref{H_CaM1}.
Upon quantization of the intra-molecular and solvent  vibrational modes, $\mathcal{H}_0$ gets identical to the Holstein-Tavis-Cummings Hamiltonian \cite{wu2016polarons,zeb2017exact,herrera_absorption_2017}.
In general, its eigenvalues and eigenstates have to be computed numerically.
In order to have analytical insight into the physics below this diagonalization, we make use of a generalized Born-Oppenheimer approximation \cite{tully_perspective_2000,galego_cavity-induced_2015}, taking into account the time-scale separation between slow nuclei motion $(\hbar\omega_{\mathrm{v}} \approx 50 \mbox{ meV})$ and the fast dynamics of strongly-coupled electrons and cavity-mode $(\Delta_{ge} \approx \hbar\omega_c \approx 2.8 \mbox{ eV})$.
We introduce the following notations for $q_{\mathrm{v},i}=Q_{\mathrm{v},i}-\overline{Q}_{\mathrm{v},g}$ and $q_{\mathrm{S},ik}=Q_{\mathrm{S},ik}-\overline{Q}_{\mathrm{S},gk}$ the displacements of the intra-molecular and solvent vibrational modes with respect to the ground-state equilibrium nuclear configuration. 
The shift of the equilibrium nuclear positions $\Delta\overline{Q}_{\mathrm{v},r}=\overline{Q}_{\mathrm{v},r}-\overline{Q}_{\mathrm{v},g}$
and $\Delta\overline{Q}_{\mathrm{S},rk}=\overline{Q}_{\mathrm{S},rk}-\overline{Q}_{\mathrm{S},gk}$ in each excited electronic state $r$ (see displaced parabolas in Fig.\ref{fig:Fig1}b), is due to electron-phonon interactions.
The corresponding electron-phonon coupling strengths are given by the reorganisation energies \cite{marcus_theory_1956,marcus_electron_1993,kestner_thermal_1974} of intra-molecular and solvent vibrations, defined respectively as $\lambda_{\mathrm{v},r}=\omega_\mathrm{v}^2\Delta\overline{Q}^2_{\mathrm{v},r}/2$ 
and $\lambda_{\mathrm{S},r}= \sum_k \lambda_{\mathrm{S},rk}$, with $\lambda_{\mathrm{S},rk}=\omega_k^2 \Delta\overline{Q}^2_{\mathrm{S},rk}/2$.
We introduce the usual adimensional Huang-Rhys factors \cite{huang2000theory} $g_{\mathrm{v},r} = \Delta\overline{Q}_{\mathrm{v},r}/2x_{0\mathrm{v}}$ and $g_{\mathrm{S},rk} = \Delta\overline{Q}_{\mathrm{S},rk}/2x_{0\mathrm{S},k}$, which are nothing but the shifts of the modes' equilibrium positions in units of the zero-point motions $x_{0\mathrm{v}}=\sqrt{\hbar/2\omega_\mathrm{v}}$ and $x_{0\mathrm{S},k}=\sqrt{\hbar/2\omega_k}$.
Huang-Rhys factors are related to reorganisation energies by the relations
$g^2_{\mathrm{v},r} = \lambda_{\mathrm{v},r}/\hbar\omega_{\mathrm{v}}$ and $g^2_{\mathrm{S},rk} = \lambda_{\mathrm{S},rk}/\hbar\omega_k$.
We proceed by first pre-diagonalizing $\mathcal{H}_0$, in the limit of vanishing electron-phonon coupling $(\lambda_{\mathrm{v},r}=\lambda_{\mathrm{S},r}=0)$. 
In this limit, nuclear motion and polariton dynamics are factorizable, so that 
an exact solution can be given for the eigenstates and eigenfunctions of $\mathcal{H}_0$
\cite{tavis_exact_1968}.
Finally, we compute analytically by perturbation theory \cite{cohen1998mecanique} the lowest non-vanishing order corrections in the electron-phonon coupling strength and add them to the zero-order terms to find approximate expressions of the PPES. 
%

\subsubsection{Ground-state}
\label{Ground_state}
The exact many-body ground-state $\ket{\mathcal{G}}$ and eigenenergy $\mathcal{E}_{\mathcal{G}}$ of $\mathcal{H}_0$ are given by 
\begin{eqnarray}
\ket{\mathcal{G}} &=& \ket{G}\otimes\ket{0_c}
\,,\label{Gs1} \\
\mathcal{E}_{\mathcal{G}} &=& \varepsilon_{\rm{G}} +
\sum_{i=1}^N \frac{\omega_{\mathrm{v}}^2}{2} q_{\mathrm{v},i}^2
+
\sum_{i=1}^N\sum_{k}\frac{\omega_k^2}{2} q_{\mathrm{S},ik}^2 
\,,\label{Gs2}
\end{eqnarray}
with $\ket{G}=\ket{g_1,\cdots,g_N}$ the product of the electronic ground-states for $N$ molecules,  
and $\ket{0_c}$ the vacuum state of the electromagnetic cavity-mode.
The ground-state PES of Eq.\ref{Gs2} is shown in Fig.\ref{fig:Fig2} (plain green curve). 
It is the sum of an electronic part $\varepsilon_{\rm{G}} = N \varepsilon_{g}+\hbar\omega_c/2$, corresponding to the energy of $N$ independent molecules in their ground-state $g$ and the cavity-mode in its vacuum ground-state, plus a quadratic contribution of vibrational oscillations around the ground-state equilibrium configurations of intramolecular and solvent modes. 
We note that the inclusion of counter-rotating terms in Eq.\ref{H_CaM4} would induce a
Lamb-shift of the ground-state energy that can be taken into account either by second-order perturbation theory \cite{mandal2020polariton}, or by full numerical diagonalization. 
Such an effect (not considered here) becomes important in the ultrastrong coupling regime, when the collective vacuum Rabi splitting is a significant portion or larger than the optical frequency $\tilde{\Omega}_R \geq \omega_c$ \cite{ciuti_quantum_2005}. 
%

\subsubsection{Upper and lower polaritons}
\label{Upper_lower_polaritons}

The RWA in Eq.\ref{H_CaM4} enables to separate the energy-sector 
corresponding to at most one cavity-photon or one molecular excitation from the higher-energy sectors and from the ground-state one.
We obtain the first upper ($\ket{\rho=+}$) and lower ($\ket{\rho=-}$) polariton manybody eigenstates
as
\begin{eqnarray}
\ket{+} &=& \cos(\theta) \ket{G}\otimes\ket{1_c}
+ \sin(\theta)\ket{E_1}\otimes\ket{0_c}
\,,\label{LUP1} \\
\ket{-} &=& -\sin(\theta) \ket{G}\otimes\ket{1_c}
+ \cos(\theta) \ket{E_1}\otimes\ket{0_c}
\,,\label{LUP2}
\end{eqnarray}
with
\begin{eqnarray}
\ket{E_1} &=& \frac{1}{\sqrt{N}}\sum_{i=1}^{N}\ket{(e_i)}
\,.\label{LUP3}
\end{eqnarray}
The coefficients in front of the manybody states in Eq.\ref{LUP1} are $\cos(\theta) = \sqrt{\alpha_{-}}$ and $\sin(\theta) = \sqrt{\alpha_{+}}$ with
\begin{eqnarray}
\alpha_{\rho=\pm} &=& \frac{1}{2}\left( 1 -\rho \frac{\delta}{\tilde{\Omega}_R} \right)
\,. \label{Alpha} 
\end{eqnarray}
The totally symmetric molecular state $\ket{E_1}$ is obtained as the sum of all states containing $N-1$ molecules in the ground-state and one molecule $i$ in the excited state $\ket{(e_i)}\equiv\ket{g_1,\cdots,g_{i-1},(e_i),g_{i+1},\cdots,g_N}$.
The electronic excitation in this $\ket{E_1}$ Dicke-state is thus delocalized on the whole molecular ensemble, the former playing the role of a giant collective dipole oscillating in phase with the electromagnetic cavity-mode \cite{dicke_coherence_1954}.
The polaritons in Eq.\ref{LUP1} and Eq.\ref{LUP2} are linear combinations of two states: one involving the manybody electronic ground-state $\ket{G}$ with one photon populating the cavity and the other the collective Dicke-state $\ket{E_1}$ with the cavity in its quantum mechanical ground-state.
The coefficients $\cos(\theta)$ and $\sin(\theta)$ are function of both the cavity-molecule detuning $\delta$ and collective vacuum Rabi splitting $\tilde{\Omega}_R$ given by
\begin{eqnarray}
\tilde{\Omega}_R = \sqrt{\delta^2 + \left( \Omega_R\sqrt{N}\right)^2}
\,.\label{LUP6}
\end{eqnarray}
As expected, $\tilde{\Omega}_R$ scales with $\sqrt{N}$, or more precisely with the 
square-root of the molecular concentration $\sqrt{N/V}$ \cite{tavis_exact_1968,haroche_cavity_nodate,houdre_early_2005}.
At resonance between the cavity-mode frequency and the molecular transition, $\delta=0$ and $\cos(\theta)=\sin(\theta)=1/\sqrt{2}$, such that the polaritons are half-matter, half-light hybrid excitations. 
At strong-detuning ($\delta \rightarrow \pm \infty$), the polariton states coincide back with the bare molecular ground and excited states. 
We obtain the PPES $\mathcal{E}_{\rho=\pm}$ corresponding to $\ket{\rho=\pm}$  
\begin{eqnarray}
&&\mathcal{E}_{\rho} = \varepsilon_{\rho} +
\sum_{i=1}^N \frac{\omega_{\mathrm{v}}^2}{2} \left(  q_{\mathrm{v},i} - \Delta \overline{Q}_{\mathrm{v},\rho}\right)^2
\, \nonumber \\
&& \qquad \quad \; +
\sum_{i=1}^N\sum_{k}\frac{\omega_k^2}{2} \left( q_{\mathrm{S},ik} - \Delta \overline{Q}_{\mathrm{S},\rho k}\right)^2
\,, \label{PPES1}
\end{eqnarray}
with $\varepsilon_{\rho}$ the polariton energy
\begin{eqnarray}
\varepsilon_{\rho} &=& \varepsilon_{\rm{G}} + \hbar\omega_c -
\frac{\hbar}{2}\left( \underline{\delta} - \rho \underline{\tilde{\Omega}}_R \right)
\,, \label{PPES4} \\
\underline{\delta} &=& \delta - \frac{\lambda_{\mathrm{v},e}+\lambda_{\mathrm{S},e}}{\hbar}
\left( 
1 - \frac{\alpha^2_+ + \alpha^2_-}{N}
\right)
\,, \label{PPES5} \\
\underline{\tilde{\Omega}}_R &=& \tilde{\Omega}_R - \frac{\lambda_{\mathrm{v},e}+\lambda_{\mathrm{S},e}}{\hbar}\frac{\delta}{\tilde{\Omega}_R}
\left( 
1 - \frac{1}{N}
\right)
\,, \label{PPES6} 
\end{eqnarray}
and $\Delta \overline{Q}_{\mathrm{v},\rho}$ and $\Delta \overline{Q}_{\mathrm{S},\rho k}$ the respective shifts in the intra-molecular and solvent modes' equilibrium positions
\begin{eqnarray}
\Delta \overline{Q}_{\mathrm{v},\rho} &=& \alpha_{\rho} \frac{\Delta \overline{Q}_{\mathrm{v},e}}{N}
\,, \label{PPES2} \\
\Delta \overline{Q}_{\mathrm{S},\rho k} &=& \alpha_{\rho}\frac{\Delta \overline{Q}_{\mathrm{S},ek}}{N}
\,. \label{PPES3} 
\end{eqnarray}
As expected, the polariton energy in Eq.\ref{PPES4} depends
on both the molecule-cavity detuning $\underline{\delta}$ and collective vacuum Rabi frequency $\underline{\tilde{\Omega}}_R$.
However, at lowest-order in the electron-phonon coupling strength, both quantities are renormalized in Eq.\ref{PPES5} and Eq.\ref{PPES6}, and become explicitly dependent on the 
intra-molecular and solvent reorganization energies (respectively $\lambda_{\mathrm{v},e}$ and $\lambda_{\mathrm{S},e}$), as well as on the number $N$ of molecules. 
At this level of approximation, the contribution of nuclear motion to the PPES in Eq.\ref{PPES1} is still quadratic, but with new equilibrium positions $\Delta \overline{Q}_{\mathrm{v},\rho}$ and $\Delta \overline{Q}_{\mathrm{S},\rho k}$ for the intra-molecular and solvent modes, both depending on detuning, collective Rabi frequency and number of molecules.
The shifts in equilibrium positions in Eq.\ref{PPES2} and Eq.\ref{PPES3} are the same for each molecule, thus corresponding to the excitation of a long-range vibrational mode, in which each molecular vibration couples in phase with the same polariton.
In the large-$N$ limit, we recover the results of the collective decoupling mechanism between nuclear motion and the polariton, as derived in Ref.\cite{herrera_cavity-controlled_2016}, for which the configuration of the nuclear equilibrium positions gets back to the ground-state configuration ($\Delta \overline{Q}_{\mathrm{v},\rho} \approx 0$ and $\Delta \overline{Q}_{\mathrm{S},\rho k} \approx 0$ when $N \gg 1$).
Eq.\ref{PPES1} is a direct physical consequence of the generalized Born-Oppenheimer approximation and perturbation expansion at the lowest-order of the electron-phonon coupling strength.
This approach generalizes previous results of Refs.\cite{herrera_cavity-controlled_2016,semenov2019electron,mandal2020polariton} by taking into account on the same footing the finite number $N$ of molecules, finite molecule-cavity detuning, and dressing of the polariton by molecular vibrations of the solvent environment. 
The PPES in Eq.\ref{PPES1} interpolates smoothly between the limits of single-molecule $N=1$ and large number of molecules $N\gg 1$ inside the cavity (terms of leading order $\approx 1/N$).
It also consistent with previous methods of approximation based on the use of the variational polaron ansatz \cite{toyozawa1954theory,silbey_electronic_1976,bera2014generalized,herrera_cavity-controlled_2016}.   
We present in Fig.\ref{fig:Fig2} the PPES for the lower polariton state $\mathcal{E}_{\rm{-}}$
(plain red curve) and upper polariton state $\mathcal{E}_{\rm{+}}$ (plain blue curve) as obtained
from Eq.\ref{PPES1}.
For comparison, the PPES obtained by numerical diagonalization of $\mathcal{H}_0$ (within RWA) are plotted in Fig.\ref{fig:Fig2} as lower and upper triangles, standing respectively for the lower and upper PPES. 
We show a very good matching of the exact numerical curves and analytical results of Eq.\ref{PPES1}, in the moderate to strong-coupling regime for which the effective Rabi frequency 
    is in the range $\hbar \omega_c > \hbar \tilde{\Omega}_R > \lambda_{\mathrm{v},e}, \lambda_{\mathrm{S},e}, \omega_{\mathrm{v}}$.
%

\subsubsection{Dark states}
\label{Dark_States}

The spectrum of $\mathcal{H}_0$ in the single-photon excitation sector, also contains a manifold of 
$N-1$ degenerate states uncoupled to the cavity-mode. 
The expression of those dark states $\ket{\mathcal{D}_{p}}$ is more complex than the one of the bright polaritons \cite{dukalski2013high,ozhigov2016space}.
It can be obtained exactly in the case of vanishing electron-phonon coupling strength
\begin{eqnarray}
\ket{\mathcal{D}_{p}} &=& \frac{1}{\sqrt{p+1}} \left( 
\frac{1}{\sqrt{p}}\sum_{j=1}^{p} \ket{\left(e_j\right)}
-
\sqrt{p} \ket{\left(e_{p+1}\right)}
\right)
\otimes\ket{0_c}
\,, \nonumber \\
\label{DS1} \\
\mathcal{E}_{\mathcal{D}_{p}} &=& \varepsilon_{\rm{D}} +
\sum_{i=1}^N
\frac{\omega_{\mathrm{v}}^2}{2} q_{\mathrm{v},i}^2
+
\sum_{i=1}^N\sum_{k}\frac{\omega_k^2}{2} q_{\mathrm{S},ik}^2 
\,,\label{DS2}
\end{eqnarray}
with $p=1,\cdots,N-1$ an index labelling the dark state, and $\varepsilon_{\rm{D}}\equiv \varepsilon_{\mathrm{G}} + \Delta_{ge}$ the dark state energy (independent of $p$).
Within RWA, those states do not couple directly to the optical cavity-mode. 
Their PES in Eq.\ref{DS2} is thus independent of the collective vacuum Rabi splitting.
In the case of finite arbitrary electron-phonon interactions, the dark PES can only be computed numerically, similarly to the Holstein polaron problem \cite{holstein1959studies}.
We obtain numerically a lifting of the dark PES degeneracy, with the creation of a miniband of states between the lower and upper polaritons.
The miniband width is proportional to the total reorganization energy $\lambda_{\mathrm{v},e}+\lambda_{\mathrm{S},e}$. 
The coupling to molecular vibrations thus broadens the manifold of dark states as does an inhomogeneous static disorder \cite{houdre_vacuum-field_1996}.
We plot in Fig.\ref{fig:Fig2} the miniband of dark PES $\mathcal{E}_{\mathcal{D}_{p}}$ obtained numerically (black dots), compared to the analytical PES given by Eq.\ref{DS2} (plain black curve). 
The former is a good approximation to the average position of the miniband. 
In the rest of the paper, we will use the analytical expression given by Eq.\ref{DS2},
even in cases for which the electron-phonon interaction is finite, which is a good approximation 
if the broadening of the miniband is smaller than the vacuum Rabi splitting. 
Finally, there are additional eigenstates of $\mathcal{H}_0$ that do not couple to the optical cavity-mode and are thus ``dark", but play an important role regarding the chemical reactivity of the confined molecules. 
Such is the case for the excited-states $\ket{(r_i)}\equiv\ket{g_1,\cdots,g_{i-1},(r_i),g_{i+1},\cdots,g_N}$ containing the molecule number $i$ in the excited electronic state $r=f$ or $r=g'$, while the remaining $N-1$ molecules are in the ground-state $g$. 
The corresponding manybody state $\ket{(\mathcal{R}_i)}$ and eigenenergy $\mathcal{E}_{\rm{\mathcal{R}_i}}$ for $r=f,\,g'$ are given by 
\begin{eqnarray}
&&\ket{\mathcal{R}_i} = \ket{\left(r_i\right)}\otimes\ket{0_c}
\,, \label{DS3} \\
&&\mathcal{E}_{\rm{\mathcal{R}_i}} = \varepsilon_{\rm{R}} +
\sum_{j=1,j\neq i}^N  
\frac{\omega_{\mathrm{v}}^2}{2} q_{\mathrm{v},j}^2
+
\sum_{j=1,j\neq i}^N \sum_{k}\frac{\omega_k^2}{2} q_{\mathrm{S},jk}^2 
\nonumber \\
&& \quad +
\frac{\omega_{\mathrm{v}}^2}{2} \left( q_{\mathrm{v},i} - \Delta \overline{Q}_{\mathrm{v},r} \right)^2
+
\sum_{k}\frac{\omega_k^2}{2} \left( q_{\mathrm{S},ik} - \Delta \overline{Q}_{\mathrm{S},rk} \right)^2 
\,, \nonumber \\
\label{DS4}
\end{eqnarray}
with $\varepsilon_{\rm{R}}\equiv \varepsilon_{\mathrm{G}} + \Delta_{gr}$ the $r$-state energy. 
The corresponding PES $\mathcal{E}_{\rm{\mathcal{R}_i}}$ are $N$-fold degenerate.
We plot $\mathcal{E}_{\rm{\mathcal{F}_i}}$ in Fig.\ref{fig:Fig2} as a dashed yellow curve.
%

\subsubsection{The concept of \textit{reacton}}
\label{Reacton}

The PPES in the subsections Sec.\ref{Upper_lower_polaritons} and Sec.\ref{Dark_States} have a simple interpretation. 
They arise from the collective dipole coupling between the electronic $g$ and $e$ states 
of the molecules and a single electromagnetic cavity-mode, resulting in the formation of a polariton.
This polariton gets further dressed by interactions with a bath of intra-molecular and solvent vibrational modes, thus sharing some similarities with the concept of polaron \cite{holstein1959studies} in solid-state physics.
The dressed polariton is however more complex than a single polaron excitation, since it involves many different energy scales \cite{hutchison_modifying_2012} ranging from molecular vibrational frequencies $\hbar\omega_{\mathrm{v}} \approx 10 \mbox{ meV}$, electronic transitions and cavity optical frequency $\Delta_{ge} \approx \hbar\omega_c \approx 2 \mbox{ eV}$, as well as the collective vacuum Rabi frequency $\hbar\tilde{\Omega}_R\approx 0.7 \mbox{ eV}$ that is intermediate between the vibronic and optical frequency scales.
We call this dressed and collective polariton excitation a \textit{reacton}, since, as we will show later, the formation of this entity modifies significantly the chemical properties of confined and resonant molecules inside the cavity. 
The concept of \textit{reacton} is a key concept that generalizes and unifies several previous investigations in the field of polaritonic chemistry \cite{galego_many-molecule_2017,herrera_cavity-controlled_2016,cwik_excitonic_2016},
and shares conceptual similarities to the \textit{dressed-atom} approach in quantum optics \cite{canaguier-durand_non-markovian_2015, cohen1998atom}.
While in this paper we compute the \textit{reacton} properties within the 
range of validity of the Born-Oppenheimer approximation \cite{galego_cavity-induced_2015}, in general, those have to be computed numerically self-consistently \cite{cwik_excitonic_2016}. 
%

\section{Charge-transfer reaction rate}
\label{Charge_transfer_reaction_rate}
In this section, we investigate the modification of chemical reactivity for cavity-confined molecules, induced by the \textit{reacton} formation.
Due to the weak but non-vanishing matrix elements ($V_{ef} \ne 0$) in the Hamiltonian $V_{\rm{CT}}$ (see Eq.\ref{H_CaM5}), molecules that are in the excited electronic state $e$ (valley of reactant) may undergo a CT process towards the other excited electronic state $f$ (valley of product), assisted by a reorganization of the molecular nuclei configuration.
The theoretical framework for describing the kinetics of such CT chemical reactions in solution was developed mainly by the works of Marcus \cite{marcus_theory_1956,marcus_electron_1993,siders_quantum_1981}, 
Kestner et al. \cite{kestner_thermal_1974}, Freed et al. \cite{freed_multiphonon_1970} and Hopfield \cite{hopfield_electron_1974}.
Our approach generalizes this framework to the case of PPES for the chemical reaction written in the \textit{reacton} basis (see Sec.\ref{PPES}), rather than in the bare molecular basis.
%
%
%
\subsubsection{Marcus theory applied to the \textit{reacton}}
\label{Marcus_Theory_Reacton}
%
\begin{figure}[!h]
\centering
\includegraphics[width=1.0\linewidth]{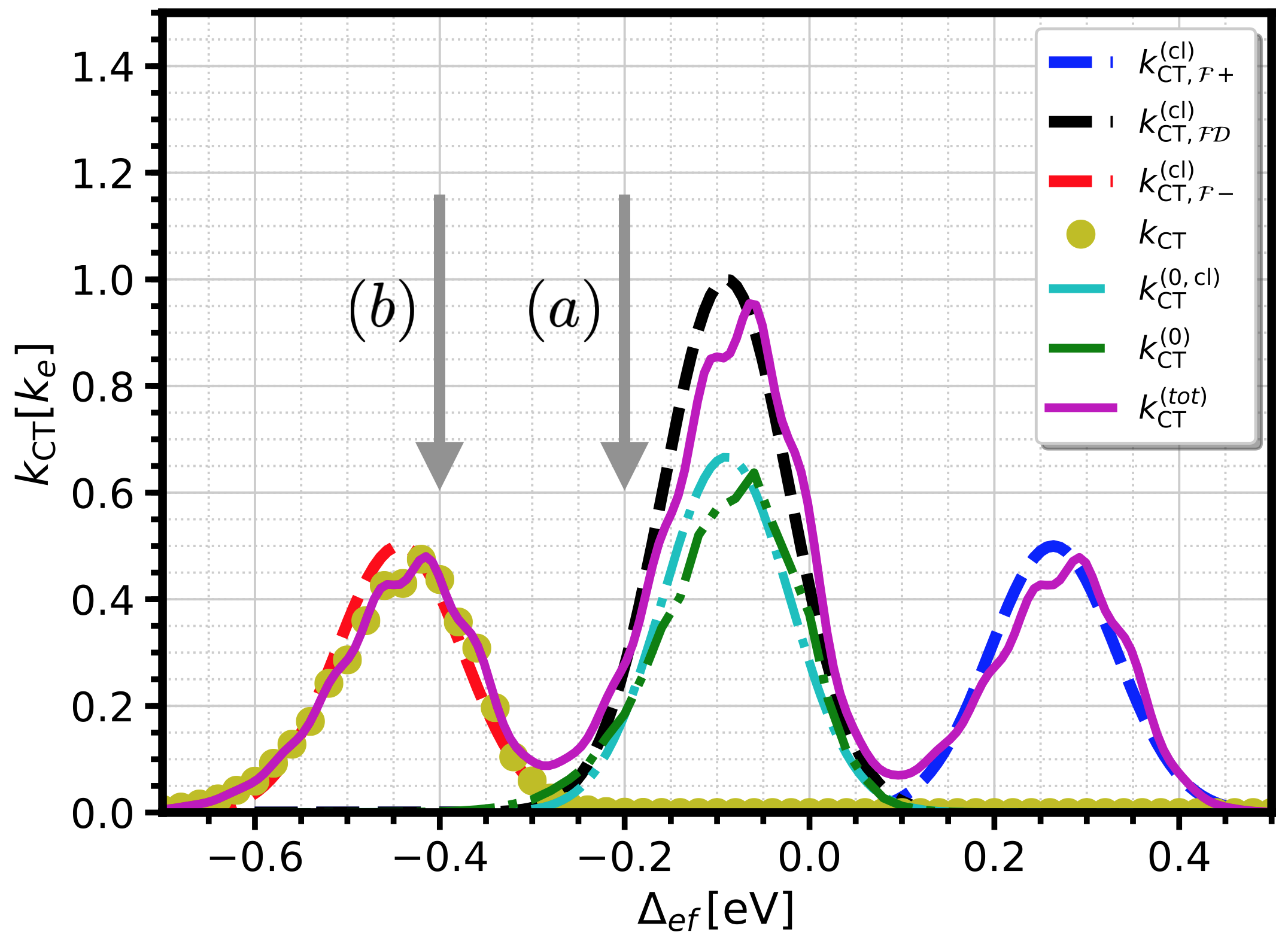}
\caption{CT thermal reaction rate inside cavity $k_{\rm{CT}}$ (yellow dotted curve) as a function of the bare reaction driving-force $\Delta_{ef}$. 
The total rate $k^{(\rm{tot})}_{\mathrm{CT}}$ is presented as a purple plain curve.
Classical contributions of the PPES to $k_{\rm{CT}}$ are shown as dashed curves for the rates $k^{(\mathrm{cl})}_{\rm{CT},\mathcal{F}-}$ (in red), $k^{(\rm{cl})}_{\rm{CT},\mathcal{F}+}$ (in blue), and $k^{(\rm{cl})}_{\mathrm{CT},\mathcal{FD}}$ (in black).
The thermal rate $k^{(0)}_{\rm{CT}}$ and classical rate $k^{(0,\mathrm{cl})}_{\rm{CT}}$  outside cavity (for $\hbar\tilde{\Omega}_R \approx 0.0 \mbox{ eV}$) are shown respectively as 
dashed-dotted green and cyan curves.
The grey $(a)$ and $(b)$ arrows are two specific values of $\Delta_{ef}$, the first one corresponding to the molecule of Fig.\ref{fig:Fig1}. 
Chosen parameters are : $N=5000$, $k_BT= 26 \mbox{ meV}$, $\varepsilon_{g}=0 \mbox{ eV}$, $\varepsilon_{e}=2.8 \mbox{ eV}$, $\varepsilon_{f}=2.6 \mbox{ eV}$, $\hbar\omega_c= 2.8 \mbox{ eV}$, $\hbar\omega_{\mathrm{v}}= 50 \mbox{ meV}$, $\hbar\omega_k = 0.1 \mbox{ meV}$, $\hbar\Omega_R= 10 \mbox{ meV}$ ($\hbar\tilde{\Omega}_R= 0.7 \mbox{ eV}$), $\hbar\delta= 0 \mbox{ eV}$, $\lambda_{\mathrm{v},e}=0.1 \mbox{ meV}$ ($\tilde{\lambda}_{\mathrm{v},\rho\mathcal{F}}= 80 \mbox{ meV}$), $\lambda_{\mathrm{S},e}=0 \mbox{ meV}$ ($\tilde{\lambda}_{\mathrm{S},\rho\mathcal{F}}= 10 \mbox{ meV}$).
}
\label{fig:Fig3}
\end{figure}
Using standard Fermi's Golden Rule, we compute the CT thermal reaction rate $k_{\rm{CT}}$ \cite{kestner_thermal_1974,siders_quantum_1981}. 
In the \textit{reacton} basis, $k_{\rm{CT}}$ is the sum
of partial contributions to the rate $k_{\mathrm{CT},\mathcal{F}\rho}$ from each PPES $\rho=\pm,\mathcal{D}$ belonging to the valley of reactant towards the valley of products $\mathcal{F}$.
This sum is ponderated by Boltzmann weights standing for thermal occupation of the valley of reactant \footnote{We generalize this approach in Sec.\ref{Ultrafast_reaction_kinetics} to cases where the occupation of the PPES are out-of-equilibrium.}
\begin{eqnarray}
k_{\rm{CT}} &=& \sum_{\rho=\pm,\mathcal{D}} \frac{e^{-\varepsilon_{\rho}/k_BT}}{Z_e} k_{\mathrm{CT},\mathcal{F}\rho} 
\,, \label{kET1} \\
k_{\mathrm{CT},\mathcal{F}\rho} &=& \alpha_{\rho}\frac{2\pi}{\hbar}|V_{ef}|^2
\mathcal{L}_{\mathrm{v},\rho\mathcal{F}} \star
\mathcal{L}_{\rm{cl}}\left(\Delta_{\rho\mathcal{F}},\tilde{\lambda}_{\mathrm{S},\rho\mathcal{F}}\right)
\,, \label{kET2} 
\end{eqnarray}
with $\Delta_{\rho\mathcal{F}}=\varepsilon_{\rm{F}}-\varepsilon_{\rho}$ the driving-force of the chemical reaction, and $\tilde{\lambda}_{\mathrm{S},\rho\mathcal{F}}$ the solvent reorganisation energies renormalized by the \textit{reacton} formation given by
\begin{eqnarray}
\tilde{\lambda}_{\mathrm{S},\pm\mathcal{F}} &=& \sum_k \hbar\omega_k\left( g_{\mathrm{S},fk} - \alpha_\pm \frac{g_{\mathrm{S},ek}}{N} \right)^2 + \alpha_\pm \frac{\lambda_{\mathrm{S},e}}{2N}\left(1-\frac{1}{N}\right)
\,, \nonumber \\
\label{Reorg1} 
\end{eqnarray}
and $\tilde{\lambda}_{\mathrm{S},\mathcal{D}\mathcal{F}} = \lambda_{\mathrm{S},f}$.
We write $Z_e$ the partition function for the reactant valley, and $\alpha_{\rho}$ the prefactors given by Eq.\ref{Alpha} for $\alpha_{\pm}$ and $\alpha_{\mathcal{D}}=1$.
Interestingly, the CT rate in Eq.\ref{kET2} is the convolution $\mathcal{L}_{\mathrm{v},\rho\mathcal{F}} \star \mathcal{L}_{\rm{cl}}\left(E,\tilde{\lambda}_{\mathrm{S},\rho\mathcal{F}}\right)\equiv\int dE' \mathcal{L}_{\mathrm{v},\rho\mathcal{F}}\left(E'\right)
\mathcal{L}_{\rm{cl}}\left(E-E',\tilde{\lambda}_{\mathrm{S},\rho\mathcal{F}}\right)$ between an intra-molecular vibrational lineshape $\mathcal{L}_{\mathrm{v},\rho\mathcal{F}}\left(E \right)$ and a solvent lineshape $\mathcal{L}_{\rm{cl}}\left(E \right)$.
As expected, the bath of solvent modes broadens the intra-molecular vibrational lineshape along the RC. 
We note here the usual separation of time scales between ``fast" intra-molecular vibrons 
$(\hbar\omega_{\mathrm{v}} \approx 50 \mbox{ meV} > k_BT)$ and ``slow" vibrational modes of the solvent
$(\hbar\omega_{k} \approx 0.1 \mbox{ meV} < k_BT)$.
This implies that in general, $\mathcal{L}_{\mathrm{v},\rho\mathcal{F}}\left(E \right)$ has to be computed considering quantum mechanical vibrational modes \cite{kestner_thermal_1974}, while $\mathcal{L}_{\rm{cl}}\left(E \right)$ is obtained in the limit of classical vibrational modes by the standard Gaussian lineshape \cite{o1953absorption,lax1952franck,kubo1955generating,marcus_electron_1993,kestner_thermal_1974}
\begin{eqnarray}
\mathcal{L}_{\mathrm{v},\rho\mathcal{F}}\left(E \right) &=&
\sum_{n_{\mathrm{v}},m_{\mathrm{v}}=0}^{+\infty} 
F_{n_{\mathrm{v}} m_{\mathrm{v}}}
\delta\left\lbrack E + \hbar\omega_{\mathrm{v}}\left( m_{\mathrm{v}} - n_{\mathrm{v}} \right) \right\rbrack
\,, \label{kET3bis} \\ 
\mathcal{L}_{\rm{cl}}\left(E,\lambda\right) &=&
\frac{1}{\sqrt{4\pi \lambda k_B T}}
\exp{\left\lbrack
-\frac{ \left( E + \lambda \right)^2}{4\lambda k_B T} \right\rbrack}
\,. \label{kET3} 
\end{eqnarray}
The coefficient $F_{n_{\mathrm{v}} m_{\mathrm{v}}}$ in Eq.\ref{kET3bis} is defined by 
\begin{eqnarray}
F_{n_{\mathrm{v}} m_{\mathrm{v}}} = e^{-g^2_{\mathrm{v},\rho\mathcal{F}}\left( 1 + 2 \overline{n}_{\mathrm{v}} \right)}
\frac{g^{2\left( n_{\mathrm{v}} + m_{\mathrm{v}} \right)}_{\mathrm{v},\rho\mathcal{F}}}{n_{\mathrm{v}}! m_{\mathrm{v}}!}
\left( 1 +  \overline{n}_{\mathrm{v}} \right)^{m_{\mathrm{v}}}
\overline{n}_{\mathrm{v}}^{n_{\mathrm{v}}}
\,, \label{kET5}
\end{eqnarray}
with $\overline{n}_{\mathrm{v}} \equiv n_{\rm{B}}\left( \hbar\omega_{\mathrm{v}} \right)$ the thermal equilibrium Bose distribution $n_{\rm{B}}\left( E \right)=\left( e^{E/k_BT} - 1\right)^{-1}$ for the intra-molecular vibrational modes.
It involves the Franck-Condon overlap $|\left\langle n_{\mathrm{v}}|\tilde{m}_{\mathrm{v}}\right\rangle|^2$ between the vibrational state $\ket{n_{\mathrm{v}}}$ belonging to the valley of reactants and the vibrational state $\ket{\tilde{m}_{\mathrm{v}}}$ belonging to the valley of products \cite{siders_quantum_1981}, the former mode being displaced by the renormalized Huang-Rhys factors
\begin{eqnarray}
g^2_{\mathrm{v},\pm\mathcal{F}} = \left( g_{\mathrm{v},f} - \alpha_\pm \frac{g_{\mathrm{v},e}}{N} \right)^2 + \alpha_\pm \frac{g^2_{\mathrm{v},e}}{2N}\left(1-\frac{1}{N}\right)
\,, \label{Reorg2}
\end{eqnarray}
and $g^2_{\mathrm{v},\mathcal{D}\mathcal{F}} = g^2_{\mathrm{v},f}$.
Using Eq.\ref{kET2}, Eq.\ref{kET3bis} and Eq.\ref{kET3}, we derive the final form for the CT thermal reaction rates
\begin{eqnarray}
k_{\rm{CT},\mathcal{F}\rho} &=& \alpha_{\rho}\frac{2\pi}{\hbar} |V_{ef}|^2 
\sum_{n_{\mathrm{v}},m_{\mathrm{v}}=0}^{+\infty} 
F_{n_{\mathrm{v}} m_{\mathrm{v}}}
\mathcal{L}_{\rm{cl}}\left(\Delta^{n_{\mathrm{v}}m_{\mathrm{v}}}_{\rho\mathcal{F}},\tilde{\lambda}_{\mathrm{S},\rho\mathcal{F}}\right)
\,, \nonumber \\
\label{kET4} 
\end{eqnarray}
with $\Delta^{n_{\mathrm{v}}m_{\mathrm{v}}}_{\rho\mathcal{F}}$ the partial driving-force of the CT reaction involving the exchange of $m_{\mathrm{v}} - n_{\mathrm{v}}$ molecular phonons
\begin{eqnarray}
\Delta^{n_{\mathrm{v}}m_{\mathrm{v}}}_{\rho\mathcal{F}} = \Delta_{\rho\mathcal{F}} + \hbar\omega_{\mathrm{v}}\left( m_{\mathrm{v}} - n_{\mathrm{v}} \right)
\,. \label{kET6}
\end{eqnarray}
Eq.\ref{kET4} is one of the main result of this paper. 
Compared to standard Marcus theory \cite{marcus_electron_1993} and previous works in polaritonic chemistry \cite{herrera_cavity-controlled_2016,semenov2019electron,mandal2020polariton}, 
we derived the CT reaction rate, taking into account the \textit{reacton} formation, which 
includes the contribution of collective PPES $\rho=\pm,\mathcal{D}$ delocalized on the whole molecular ensemble, that are available to the chemical reaction. 
We notice that due to the collective nature of the \textit{reacton}, not only the reaction driving force strength $\Delta_{\rho\mathcal{F}}$ is modified (see Eq.\ref{kET6}), but also the intra-molecular vibrational Huang-Rhys factors $g^2_{\mathrm{v},\rho\mathcal{F}}$ (see Eq.\ref{Reorg2}) and solvent reorganisation energies $\tilde{\lambda}_{\mathrm{S},\rho\mathcal{F}}$ (see Eq.\ref{Reorg1}).
Finally, in the limit of ``slow" intra-molecular vibrational mode $\omega_{\mathrm{v}}<k_BT/\hbar$, Eq.\ref{kET1} formally recovers the ``semi-classical" approximation derived by Marcus \cite{siders_quantum_1981}.
In this limit, we obtain the classical CT thermal rate $k^{(\rm{cl})}_{\rm{CT}}$
\begin{eqnarray}
k^{(\rm{cl})}_{\rm{CT}} &=& \sum_{\rho=\pm,\mathcal{D}} \frac{e^{-\varepsilon_{\rho}/k_BT}}{Z_e} k^{(\rm{cl})}_{\rm{CT},\mathcal{F}\rho} \,, \label{kET7} \\
k^{(\rm{cl})}_{\rm{CT},\mathcal{F}\rho} &=& \alpha_{\rho}\frac{2\pi}{\hbar} |V_{ef}|^2
\mathcal{L}_{\rm{cl}}\left( \Delta_{\rho\mathcal{F}},\tilde{\Lambda}_{\rho\mathcal{F}}\right)
\,,\label{kET8} 
\end{eqnarray}
with total reorganization energy $\tilde{\Lambda}_{\rho\mathcal{F}}=\tilde{\lambda}_{\mathrm{v},\rho\mathcal{F}}+\tilde{\lambda}_{\mathrm{S},\rho\mathcal{F}}$.
%

\subsubsection{CT reaction rate inside the cavity}
\label{Reaction_Rate_Inside_Cavity}
%
%
\begin{figure}[!h]
\centering
\includegraphics[width=1.0\linewidth]{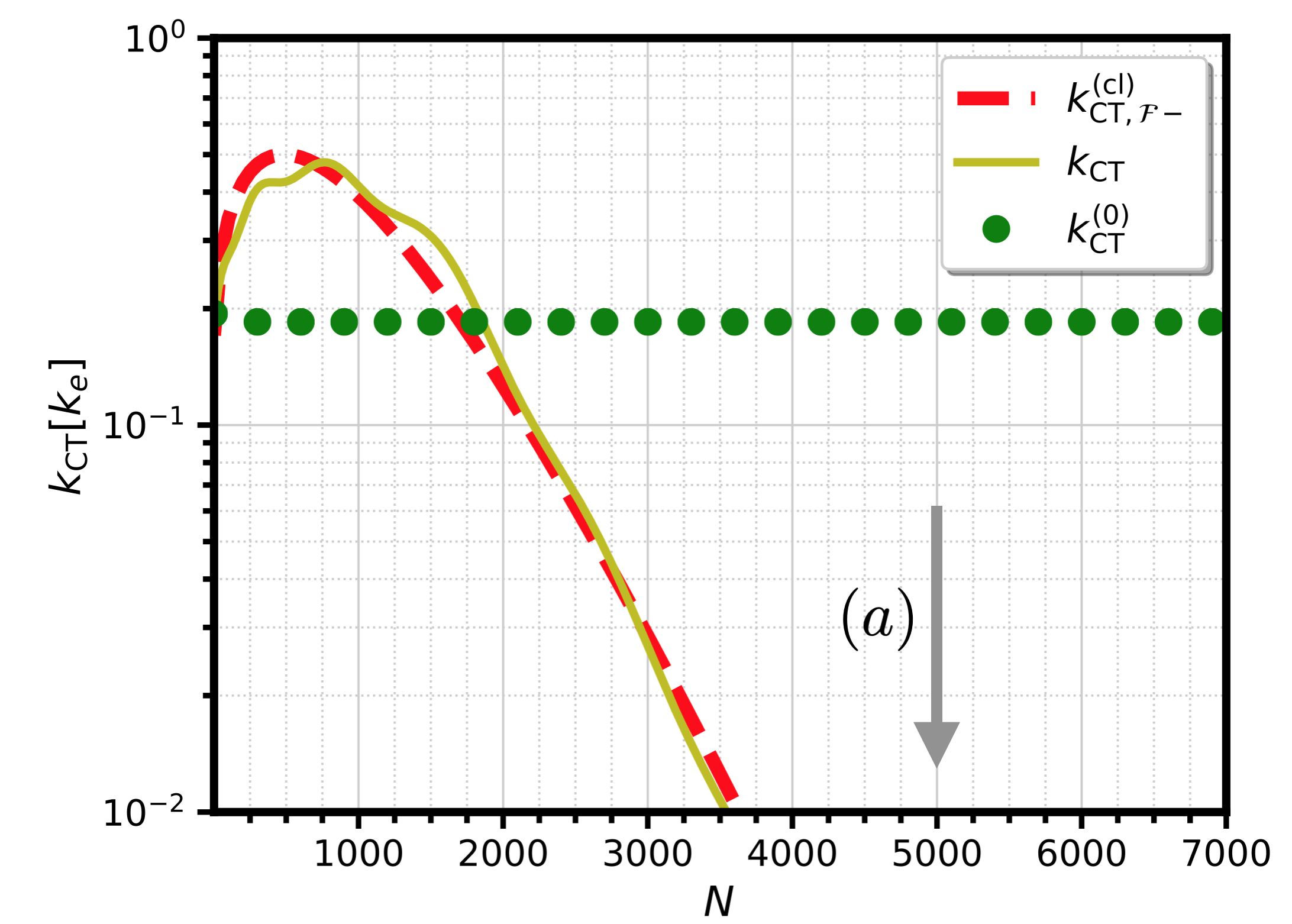}
	\caption{
	Thermal reaction rate $k_{\rm{CT}}$ (plain yellow curve) and partial classical reaction rate $k^{(\rm{cl})}_{\mathrm{CT},\mathcal{F}-}$ (dashed red curve) as a function of the number of coupled molecules $N$. 
	The thermal rate out of cavity $k^{(0)}_{\rm{CT}}$ is shown as a green dotted curve. 
	Parameters are those of Fig.\ref{fig:Fig3}, except for $N$, with the reaction driving-force
	fixed to the value $\Delta_{ef}=-0.2 \mbox{ eV}$.
	The case $N=5000$ is shown by the grey arrow ($a$) as in Fig.\ref{fig:Fig3}. 
    }
\label{fig:Fig4}
\end{figure}
%

%
%
\begin{figure}[!h]
\centering
\includegraphics[width=1.0\linewidth]{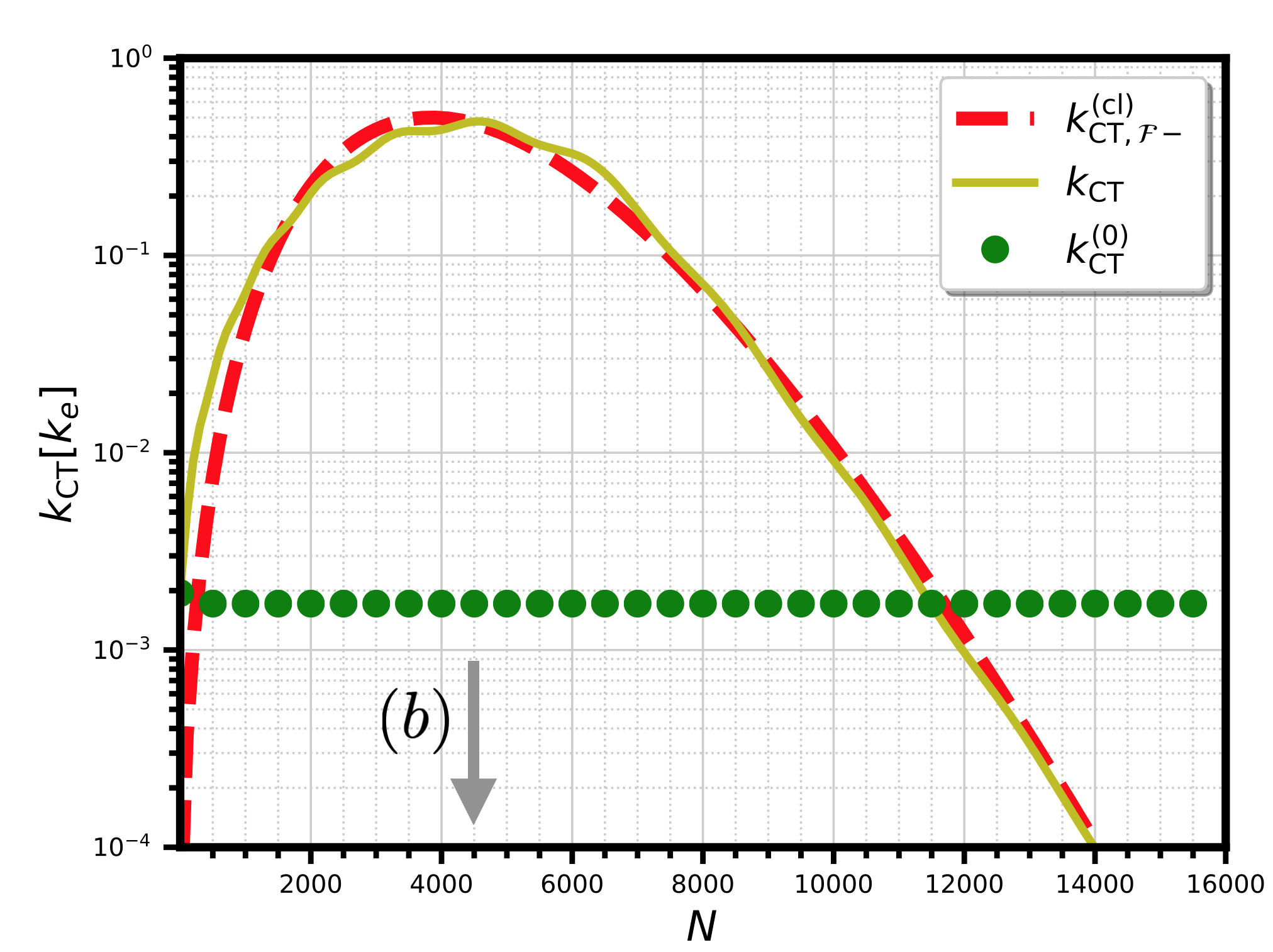}
	\caption{
	Same plot as in Fig.\ref{fig:Fig4}, but with the reaction driving-force
	fixed to the value $\Delta_{ef}=-0.4 \mbox{ eV}$.
	The case $N=5000$ is shown by the grey arrow ($b$) as in Fig.\ref{fig:Fig3}. %
    }
	\label{fig:Fig5}
\end{figure}
In the following, we focus on the case of room temperature $k_B T=26 \mbox{ meV}$ and a cavity frequency $\omega_c$ that is resonant ($\delta=0$) with the molecular transition $\Delta_{ge}/\hbar = 2.8 \mbox{ eV}/\hbar$ in Fig.\ref{fig:Fig1}b.
For a typical Fabry-P\'{e}rot cavity of surface $10^4 \mbox{ }\mu m^2$ with distant mirrors of the fundamental optical cavity-mode wavelength $\lambda_c/2 \approx \pi c/\omega_c \approx 0.221 \mbox{ }\mu m$ (for $n\approx 1$), and for molecules of electric dipole moment $\mu \approx 5 \mbox{ D}$, we estimate a very weak bare vacuum Rabi-splitting $\hbar\Omega_R \approx 0.35 \mbox{ }\mu \mbox{eV}$.
In best cases for which the molecules are in average packed $25\,\AA$ away one from each other and equally coupled to the cavity mode, we estimate the maximum number of embedded molecules $N \approx 10^{11}$ thus leading to an upper-bound for the collective vacuum Rabi-splitting of about $\hbar\tilde{\Omega}_R = 0.11 \mbox{ eV}$.
The former value is consistent with reported experimental values of $\tilde{\Omega}_R$ in nanofluidic Fabry-P\'{e}rot cavities \cite{bahsoun_electronic_2018}. 
For simplicity and illustrative purposes, we adopt a much larger value of the bare vacuum Rabi-splitting $\Omega_R = 10 \mbox{ meV}$ that is consistent with the highest single-molecule-cavity couplings $(\approx 100 \mbox{ meV})$ reported in plasmonic cavities \cite{chikkaraddy2016single}. 
We consider a population of $N=5000$ molecules coherently coupled to the same optical cavity mode, 
for which the collective vacuum Rabi-splitting $\tilde{\Omega}_R = 0.7 \mbox{ eV}$ is close to reported experimental values in optical microcavities \cite{hutchison_modifying_2012}. 
Finally, we choose the frequency of intra-molecular vibrational modes $\omega_{\mathrm{v}} \approx 50 \mbox{ meV}$ and solvent ones $\omega_k \approx 0.1 \mbox{ meV}$. 
The dressed reorganization energies are fixed to $\tilde{\lambda}_{\mathrm{v},\rho\mathcal{F}} = 80 \mbox{ meV}$ and $\tilde{\lambda}_{\mathrm{S},\rho\mathcal{F}} = 10 \mbox{ meV}$ leading to a total reorganization energy $\tilde{\Lambda}_{\rho\mathcal{F}}=90\mbox{ meV}$.
The former value corresponds to a solvent that is sufficiently apolar \cite{leontyev2005reorganization} not to screen too much electric interactions in solution but is still sufficiently polar to increase the impact of solvent fluctuations on the kinetics of the CT reaction.
We present in Fig.\ref{fig:Fig3}, the evolution of the CT thermal reaction rate $k_{\rm{CT}}$ (yellow dotted curve) computed from Eq.\ref{kET1} in units of
\begin{eqnarray}
k_{e}\equiv\frac{2\pi}{\hbar} |V_{ef}|^2/\sqrt{4 \pi \tilde{\Lambda}_{-\mathcal{F}}k_B T}
\,,\label{ke_init} 
\end{eqnarray}
as a function of the bare reaction driving-force $\Delta_{ef}\equiv \varepsilon_f - \varepsilon_e$, at fixed $\tilde{\Omega}_R = 0.7 \mbox{ eV}$.
For comparison, we plotted the total rate $k^{(\rm{tot})}_{\rm{CT}}=\sum_{\rho=\pm,\mathcal{D}} k_{\rm{CT},\mathcal{F}\rho} $ (purple plain curve), which is the sum of contributions of each PPES to the reaction rate. 
The partial and classical CT rates $k^{(\rm{cl})}_{\rm{CT},\mathcal{F}\rho}$ given by Eq.\ref{kET8} are also plotted as dashed curves.
We find that the contribution of dark states $k^{(\rm{cl})}_{\rm{CT},\mathcal{FD}}$ (in black) dominates over the two polariton satellite peaks of half amplitudes $k^{(\rm{cl})}_{\rm{CT},\mathcal{F}-}$ (in red) and $k^{(\rm{cl})}_{\rm{CT},\mathcal{F}+}$ (in blue).
The former are strongly dependent on both the detuning $\delta$ and collective vacuum Rabi frequency $\tilde{\Omega}_R$. 
They are given by two Gaussian satellite peaks centered on  
$\Delta_{ef} \approx - \tilde{\Lambda}_{\pm\mathcal{F}} + \left( \lambda_{\mathrm{v},e} + \lambda_{\mathrm{S},e} \pm \hbar\tilde{\Omega}_R \right)/2$, thus $\approx\pm 350\mbox{ meV}$ away from the main dark state peak.
The standard deviation of those curves is $\approx\sqrt{2\tilde{\Lambda}_{\pm\mathcal{F}} k_BT}$, corresponding to a full width at half maximum (FWHM) of $\approx 161 \mbox{ meV}$.
We remark that the actual CT thermal rate $k_{\rm{CT}}$ is very well approximated by the classical contribution of the lower polariton $k^{(\rm{cl})}_{\rm{CT},\mathcal{F}-}$.
On one side, this is due to the fact that $\hbar \tilde{\Omega}_R \gg k_B T$, so that only the 
lowest-energy PPES channel is significantly populated at thermal equilibrium and is thus open for the ET  reaction: the other channels $k^{(\rm{cl})}_{\rm{CT},\mathcal{FD}}$ and $k^{(\rm{cl})}_{\rm{CT},\mathcal{F}+}$ are far away in energy and thus do not contribute significantly to $k_{\rm{CT}}$ \footnote{This can be different in out-of-equilibrium situations, like the one of Sec.\ref{Ultrafast_reaction_kinetics}.}.
On the other side, we are not expecting a priori the classical approximation in Eq.\ref{kET8} to hold, since for our range of parameters, the intra-molecular vibrational modes are quantum mechanically frozen $(k_B T < \hbar\omega_{\mathrm{v}})$.
Departures from the Gaussian limit are indeed seen on the numerical plots, that manifest as the appearance of weak vibrational sidebands and asymmetries in the tails of the $k_{\rm{CT}}(\Delta_{ef})$ curve.
The former features are partially smeared out by convolution of the intra-molecular lineshape by the solvent lineshape in  Eq.\ref{kET2}, thus explaining the unexpected good qualitative match of the CT rate with the classical limit (see also Ref.\cite{siders_quantum_1981}). 
%

\subsubsection{Tuning the CT reaction rate}
\label{Tuning_ET_Rate}
\begin{table}[h!]
\centering
\footnotesize
\begin{tabular}{c|c c c c c c c} 
 \hline\hline
Rates & 
\colorbox{orange}{$k_{\mathcal{G}\mathcal{G}^{\prime}}$} &
\colorbox{orange}{$k_{\mathrm{CT},-\mathcal{F}}$} &
\colorbox{orange}{$k_{\mathrm{CT},\mathcal{FD}}$} & \colorbox{orange}{$k_{\mathrm{CT},\mathcal{F}+}$} \\ [0.5ex] 
 \hline
 meV & 3.7 & 41.4 & 42.2 & 0.001 \\ 
 THz & 0.9 & 10 & 10.2 & 0.0003  \\ [1ex] 
 \hline
\end{tabular}
\caption{Table of computed and dominant thermal reaction rates. 
The parameters are those of Fig.\ref{fig:Fig3}, for $\Delta_{ef} = -0.2 \mbox{ eV}$ (see grey arrow $(a)$).
}
\label{table:Table1}
\end{table}
To complete the picture of the reaction kinetics, we show on Fig.\ref{fig:Fig4} (plain yellow curve), the CT thermal rate $k_{\rm{CT}}$ inside cavity (with $\Omega_R=10\mbox{ meV}$) and the same rate $k^{(0)}_{\rm{CT}}$ outside cavity (for which $\Omega_R=0.0\mbox{ meV}$), as a function of the number $N$ of molecules coupled to the cavity-mode. 
The parameters are those of Fig.\ref{fig:Fig3}, with the reaction driving-force fixed at $\Delta_{ef}\approx -0.2 \mbox{ eV}$.
This choice of $\Delta_{ef}$ corresponds to the PES for the chosen molecule in Fig.\ref{fig:Fig1}b.
For the case $(a)$ labelled by a grey arrow and corresponding to $N=5000$ and $\Delta_{ef} = -0.2 \mbox{ eV}$, we find in both Fig.\ref{fig:Fig3} and Fig.\ref{fig:Fig4} that $k_{\rm{CT}} \ll k^{(0)}_{\rm{CT}}$, so that the reaction kinetics gets much slower inside than outside cavity.
Interestingly in Fig.\ref{fig:Fig4}, the CT rate does not evolves in a monotonous fashion with $N$. 
It first increases with $N$, reaching a maximum at $N \approx 500$ for which $k_{\rm{CT}} > k^{(0)}_{\rm{CT}}$ and finally slows down to $0$ with $k_{\rm{CT}} \ll k^{(0)}_{\rm{CT}}$ at large $N$.
There is thus an optimal value of $N$ (and thus of molecular concentration $N/V$ for the coupled molecules) for which the effect of vacuum quantum fluctuations of the cavity mode is maximum. 
We interpret this behavior by the modulation of the reaction driving-force $\Delta_{\rho\mathcal{F}}$ with the collective vacuum Rabi splitting $\tilde{\Omega}_R \approx \Omega_R \sqrt{N}$.
The maximum of $k^{(\rm{cl})}_{\rm{CT},\mathcal{F}-}$ in Eq.\ref{kET8} is obtained 
at the transition point to the inverted Marcus region, as is shown on Fig.\ref{fig:Fig4} (dashed red curve).
This optimal sensitivity of the CT reaction rate close to the inverted region of Marcus parabola, is in contrast to Ref.\cite{herrera_cavity-controlled_2016} that reported a monotonous increase of the reaction rate with $N$ in the resonant nuclear tunneling regime.
We provide on Table \ref{table:Table1} typical values for the cavity-induced CT reaction rates $k_{\rm{CT},\mathcal{F}\rho}$ associated to case $(a)$.
Furthermore, we estimate the reaction rate $k_{\mathcal{G'}\mathcal{G}}$ from the manybody ground-state $\mathcal{G}$ to the other manybody state $\mathcal{G}'$, using transition-state theory \cite{evans_applications_1935,wigner_transition_1938,kramers_brownian_1940,eyring_activated_1935}
\begin{eqnarray}
k_{\mathcal{G}'\mathcal{G}} &=& k_0 e^{-\frac{\Delta_{\mathcal{G}\mathrm{TS}}}{k_BT}}
\,,\label{TST1}
\label{arrh}
\end{eqnarray}
with the energy barrier $\Delta_{\mathcal{G}\mathrm{TS}}=\varepsilon_{\mathrm{TS}}-\varepsilon_{\mathcal{G}}$ between the ground-state and the transition-state, and the typical reaction rate $k_0 \approx k_B T/2\pi\hbar$.
Finally, for completeness, we show on Fig.\ref{fig:Fig5} the evolution of the CT thermal rate $k_{\rm{CT}}$ inside cavity as a function of number $N$ of coupled molecules, but for a different value of reaction driving-force fixed at $\Delta_{ef}\approx -0.4 \mbox{ eV}$.
For case $(b)$ shown as a grey arrow, for which $N=5000$ and $\Delta_{ef} = -0.4 \mbox{ eV}$, the kinetics of the CT reaction is much faster inside than outside cavity ($k_{\rm{CT}} \gg k^{(0)}_{\rm{CT}})$ in both Fig.\ref{fig:Fig3} and Fig.\ref{fig:Fig5}.
We find a similar trend as in Fig.\ref{fig:Fig4}, with a non-monotonous evolution of the CT rate with $N$.
It is thus interesting to notice that depending on the range of parameters (reaction driving-force, number of molecules, detuning), the reaction kinetics can be either slowed down or accelerated significantly by interaction with the cavity mode. 
%

\section{Dissipation}
\label{Dissipation}
%
\subsubsection{Microscopic model for dissipation}
\label{Microscopic_model_dissipation}

\begin{figure}[tbh]
\includegraphics[width=0.8\linewidth]{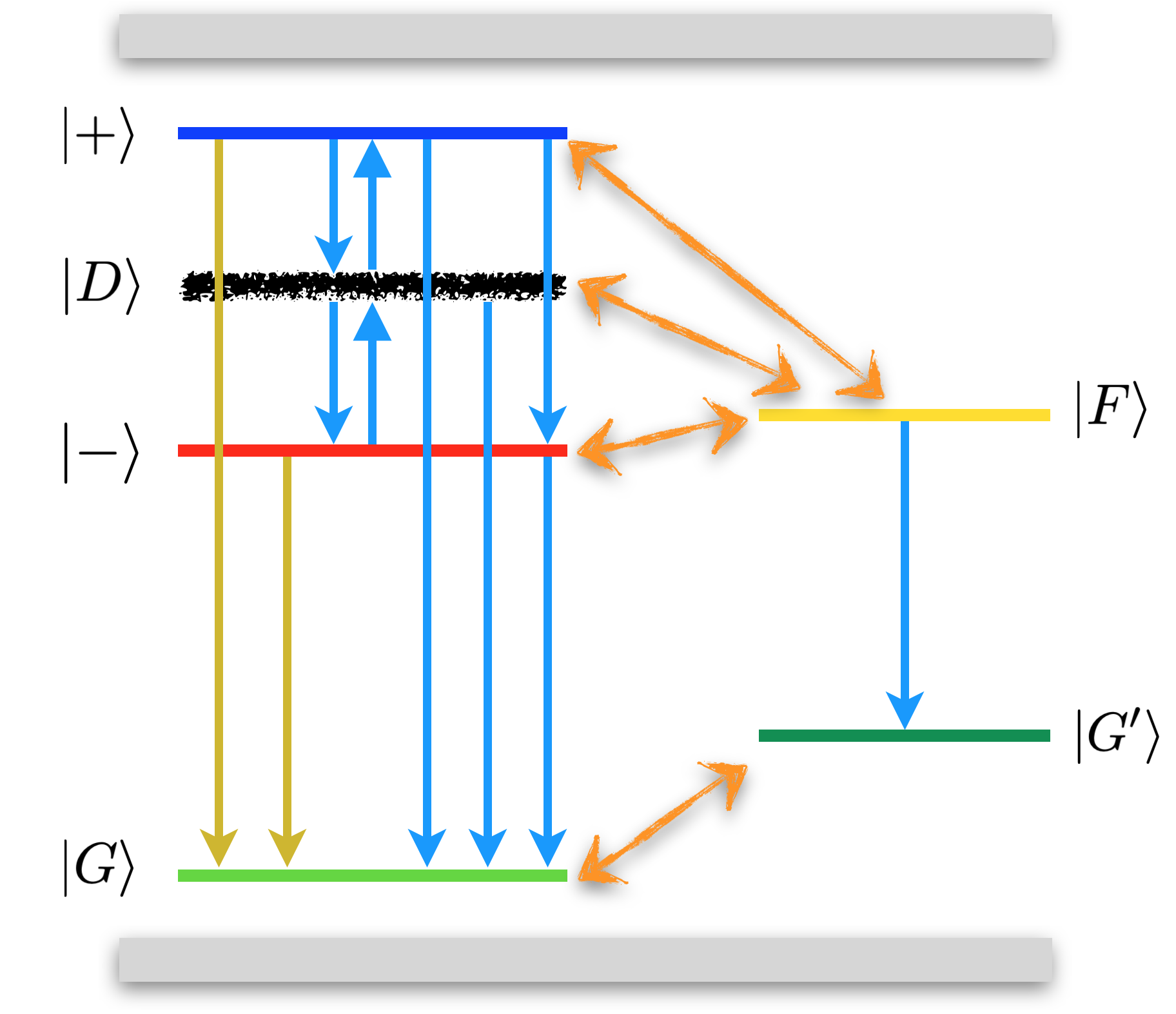}
\caption{Schematics of dissipation and dephasing rates originating from interaction between the \textit{reacton} states and the external environment.
Radiative relaxation rates are presented as gold arrows, while non-radiative relaxation and dephasing rates are both respectively shown with light-blue arrows.
The reaction rates involved in the photochemical reaction are pictured with orange double arrows.
}
\label{fig:Fig6}
\end{figure}
In this section, we introduce a minimal microscopic model of dissipation and dephasing, induced by coupling of the \textit{reacton} states to the external environment (see Fig.\ref{fig:Fig6}).
We consider two main external environments, namely the electromagnetic environment (EM) of the cavity-mode described by the Hamiltonian $H_{\rm{EM}}$ in Eq.\ref{H_Diss2}, and the solvent vibrational environment (ph) modelled by the Hamiltonian $H_{\rm{ph}}$ in Eq.\ref{H_Diss3}. 
We denote $V_{\rm{Ca-EM}}$ (in Eq.\ref{H_Diss4}) the interaction Hamiltonian between the cavity-mode and the external EM environment at the origin of photon-losses out of the cavity \footnote{We did not take into account terms at the origin of spontaneous emission in Eq.\ref{H_Diss1}, since the former occurs on nanosecond time scale while we investigate here the picosecond relaxation dynamics of the \textit{reacton}. Including spontaneous emission to our model would be straightforward.}, and $V_{\rm{M-ph}}$ (in Eq.\ref{H_Diss5}) the general Hamiltonian describing coupling between the solvated molecules and the vibrational modes of the solvent. 
The total Hamiltonian $\mathcal{H}_{\mathcal{R}-\rm{env}}$ describing the external bath environments (env) and their coupling to the \textit{reacton} ($\mathcal{R}$) is given by  
\begin{eqnarray}
&&\mathcal{H}_{\mathcal{R}-\rm{env}} = H_{\rm{EM}} + H_{\rm{ph}} + V_{\rm{Ca-EM}} +  V_{\rm{M-ph}}
\label{H_Diss1}\,,\\
&&H_{\rm{EM}} = \sum_{q} \hbar\omega_q a^\dagger_q a_q
\,,\label{H_Diss2} \\
&&H_{\rm{ph}} = \sum_{i=1}^{N} \sum_{k} \hbar\omega_k b^\dagger_{ik} b_{ik}
\,,\label{H_Diss3} \\
&&V_{\rm{Ca-EM}} = i\hbar\sum_{q} \left( f_q a_q^\dagger a -
f^*_q a^\dagger a_q \right)
\,,\label{H_Diss4} \\
&&V_{\rm{M-ph}} = \sum_{i=1}^{N} \sum_{k} \left( b_{ik} + b^\dagger_{ik} \right) \lbrace
\lambda_{e,ik} \ket{e_i} \bra{e_i}
\nonumber \\
&& +
\lambda_{ge,ik} \left( \ket{g_i} \bra{e_i} + \ket{e_i} \bra{g_i} \right)
\rbrace
\,,\label{H_Diss5} 
\end{eqnarray}
with $\omega_q$ and $\omega_k$, the respective frequencies of the electromagnetic and vibrational modes of the baths. 
$a^\dagger_q$ is the creation operator for a photon in the external EM mode with momentum $q$, while $b^\dagger_{ik}$ the creation operator for a vibron in the solvent bath associated to molecule $i$ with quasi-momentum $k$.
In Eq.\ref{H_Diss4}, $f_q$ is the probability amplitude for a cavity-photon to tunnel out 
of the cavity to the EM bath \cite{gardiner1985input,ciuti2006input}.
The electron-phonon interactions in Eq.\ref{H_Diss5} couple the quantized phonon displacement operators $ b_{ik} + b^\dagger_{ik}$ to both the electronic density of the excited state $e$ of molecule $i$ with amplitude $\lambda_{e,ik}$ (Holstein-like term \cite{holstein1959studies}) and to the off-diagonal hopping terms between states $e$ and $g$ with amplitude $\lambda_{ge,ik}$ (Su-Schrieffer-Heeger-like terms \cite{su1979solitons}).
We note that the bare PES given by Eq.\ref{H_CaM3} in Sec.\ref{Microscopic_Hamiltonian} arises (before second quantization) from electron-phonon interactions similar to the ones described by the  Holstein-like terms of Eq.\ref{H_Diss5}.
There seems thus to be a redundancy in the writing of $V_{\rm{M-ph}}$. 
However, this is not the case, since the manybody \textit{reacton} wavefunctions derived in Sec.\ref{Upper_lower_polaritons} and Sec.\ref{Dark_States} are not the exact eigenstates of the  Hamiltonian $\mathcal{H}$ (in Eq.\ref{H_CaM1}), but only approximate ones. 
Moreover, Eq.\ref{H_CaM3} doesn't contain off-diagonal coupling terms 
which are present in Eq.\ref{H_Diss5} and induce contributions to the vibrational relaxation rates.
%

%
\subsubsection{Radiative relaxation}
\label{Radiative_relaxation}
\begin{table}[h!]
\centering
\footnotesize
\begin{tabular}{c|c c} 
 \hline\hline
Rates &  \colorbox{oldgold}{$\Gamma_{\mathcal{G}-}$} & \colorbox{oldgold}{$\Gamma_{\mathcal{G}+}$} \\ [0.5ex] 
 \hline
 meV & 28 & 28 \\ 
 THz & 6.8 & 6.8 \\ [1ex] 
 \hline
\end{tabular}
\caption{Table of computed radiative relaxation rates due to cavity losses.
Parameters: same as in Fig.\ref{fig:Fig3}, for $\Delta_{ef} = -0.2 \mbox{ eV}$ (see grey arrow $(a)$).
The cavity quality factor is $Q=50$, which corresponds to a bare cavity damping rate $\kappa\approx 56 \mbox{ meV}$.
}
\label{table:Table2}
\end{table}
We consider the interaction Hamiltonian $V_{\rm{Ca-EM}}$ as a perturbation to the Hamiltonian $\mathcal{H} + H_{\rm{EM}} + H_{\rm{ph}}$ (see Eq.\ref{H_CaM1} and Eq.\ref{H_Diss1}).
%
We use Fermi's Golden Rule to compute the radiative relaxation rate $\Gamma_{\mathcal{G}\rho}$ from the manybody PPES state $\rho=\pm$ to the manybody ground-state $\mathcal{G}$ induced by $V_{\rm{Ca-EM}}$
(see Fig.\ref{fig:Fig6}, downward gold arrows).
We obtain
\begin{eqnarray}
\Gamma_{\mathcal{G}\rho} = \alpha_{-\rho}
\int dE \kappa \left(\frac{E}{\hbar}\right) J^{(em)} \left( E \right) 
\mathcal{L}_{\mathrm{ph},\mathcal{G}\rho}\left(E-\Delta_{\mathcal{G}\rho}\right)
\,, \nonumber \\
\label{Rad_Rate_1}  
\end{eqnarray}
with $\Delta_{\mathcal{G}\rho} = \varepsilon_{\rho} - \varepsilon_{\mathrm{G}}$ 
and $\kappa(\omega=E/\hbar)= 2\pi |f(\omega)|^2 \nu_{\rm{EM}}(\omega)$
the energy-dependent radiative dissipation rate of the cavity, 
given by the product of  the matrix-element square $|f_{q}|^2$ evaluated at energy $\hbar\omega_q\equiv\hbar\omega$, and the density of states of the external electromagnetic bath $\nu_{\rm{EM}}(\omega)=\sum_q \delta\left( \omega - \omega_q \right)$.
The factor $J^{(em)} \left( E \right) = 1 + n_{\rm{B}}(E)$ is associated to the emission (em) process of a photon into the electromagnetic environment that assists the downward transition.
The decay rate $\Gamma_{\mathcal{G}\rho}$ is the convolution between the cavity spectral distribution  $\kappa(E/\hbar)J^{(em)} \left( E \right)$ and the generalized vibrational lineshape $\mathcal{L}_{\mathrm{ph},\mathcal{G}\rho}\left(E\right)\equiv \mathcal{L}_{\mathrm{v},\rho\mathcal{F}} \star \mathcal{L}_{\rm{cl}}\left(E,\tilde{\lambda}_{\mathrm{S},\rho\mathcal{F}}\right)$ obtained in Sec.\ref{Marcus_Theory_Reacton}.

Eq.\ref{Rad_Rate_1} is a generalization of Refs.\cite{canaguier-durand_non-markovian_2015,pino_quantum_2015,Martinez_2018} to the case of the manybody \textit{reacton} states.
We now use the simplified assumptions that i) the energy-dependent vibrational lineshape $\mathcal{L}_{\rm{ph},\mathcal{G}\rho}\left( E \right)$ is thinner than the cavity lineshape $\kappa\left( E/\hbar \right)$, such that $\Gamma_{\mathcal{G}\rho} \approx \alpha_{-\rho} J^{(em)} \left( \Delta_{\mathcal{G}\rho}\right) \kappa \left(\Delta_{\mathcal{G}\rho}\right)$, and ii) the energy dependence of $\kappa(\omega)\approx \kappa(\omega_c) \equiv \kappa$ can be neglected on the scale of the energy difference $\Delta_{\mathcal{G}\rho}$ for the considered radiative transition (Markovian assumption), such that
\begin{eqnarray}
\Gamma_{\mathcal{G}\rho} \approx \alpha_{-\rho} J^{(em)} \left( \Delta_{\mathcal{G}\rho} \right) \kappa
\,. \label{Rad_Rate_2}
\end{eqnarray}
Within assumptions i) and ii), we obtain the corresponding upward transition rates $\Gamma_{\rho\mathcal{G}}$ from the ground-state $\mathcal{G}$ to the polariton state $\rho=\pm$ as
\begin{eqnarray}
\Gamma_{\rho\mathcal{G}} &\approx& \alpha_{-\rho} J^{(abs)} \left( \Delta_{\mathcal{G}\rho} \right) \kappa
\,, \label{Rad_Rate_3}   
\end{eqnarray}
with $J^{(abs)} \left( E \right) = n_{\rm{B}}(E)$ associated to the absorption (abs) process of a photon of the electromagnetic environment during the upward transition.
We notice however, that for the cavity mode $\hbar\omega_c \gg k_B T$ at room temperature, such that in practice $n_{\rm{B}}\left( \Delta_{\mathcal{G}\rho} \right) \ll 1$ and
$\Gamma_{\mathcal{G}\rho} \approx  \alpha_{-\rho}\kappa \gg \Gamma_{\rho\mathcal{G}} \approx 0$.
We note that relaxing assumption ii) keeping assumption i) valid, one recovers the non-Markovian calculation for the radiative relaxation made in Ref.\cite{canaguier-durand_non-markovian_2015}, that is postulated to be at the origin of the observed much shorter lifetime for the upper polariton compared to the lower one.
In the following, we will make use of both approximations i) and ii), since those are the ones that minimize the knowledge about the microscopic damping mechanism. 
Generalization to Eq.\ref{Rad_Rate_1} is possible if additional information about the energy-dependence of both optical cavity and vibrational lineshapes become available from experiments.
We estimate in Table.\ref{table:Table2} the values of typical radiative relaxation rates $\Gamma_{\mathcal{G}\rho}$ written in the \textit{reacton} basis (downward gold arrows in Fig.\ref{fig:Fig6}), from the knowledge of the bare cavity damping rate $\kappa$ and optical-cavity quality factor $Q$ in experiments\cite{schwartz2013polariton,wang2014quantum,canaguier-durand_non-markovian_2015,bahsoun_electronic_2018}.
%

\subsubsection{Non-radiative relaxation}
\label{Non_Radiative_relaxation}
\begin{table}[h!]
\centering
\footnotesize
\begin{tabular}{c|c c c c} 
 \hline\hline
Rates & \colorbox{richelectricblue}{$\gamma_{\mathcal{G^{\prime}}\mathcal{F}}$} & \colorbox{richelectricblue}{$\gamma_{\mathcal{G}\pm}$} &  \colorbox{richelectricblue}{$\gamma_{\mathcal{G}\mathcal{D}}$} &  \colorbox{richelectricblue}{$\gamma_{\mathcal{D}+}$} \\ [0.5ex] 
 \hline
 meV & 6.6 & 3 & 6 & 41.3 \\ 
 THz & 1.6 & 0.7 & 1.4 & 10 \\ [1ex] 
 \hline
\end{tabular}
\caption{Table of computed and dominant non-radiative relaxation rates due to electron-phonon interactions.
The parameters are those of Fig.\ref{fig:Fig3}, for $\Delta_{ef} = -0.2 \mbox{ eV}$ (see grey arrow $(a)$).
The bare vibronic relaxation rate is $\gamma_{\mathrm{v}}\approx 6 \mbox{ meV}$ and the dephasing rate is chosen to be $\gamma_\phi\approx 82.7 \mbox{ meV}$.
}
\label{table:Table3}
\end{table}
We compute the non-radiative relaxation rates induced by the SSH-like contributions to the electron-phonon interaction Hamiltonian $V_{\rm{M-ph}}$ in Eq.\ref{H_Diss5}.
We suppose for simplicity that the off-diagonal matrix elements $\lambda_{ge,ik}\equiv \lambda_{ge} (\omega_k)$ are independent of the molecular index.
Using similar approximations i) and ii) as in Sec.\ref{Radiative_relaxation}, we obtain for the 
dominant downwards rates
\begin{eqnarray}
\gamma_{\mathcal{G}\rho} &\approx& \alpha_{\rho} \gamma_{\mathrm{v}} \,, \label{Non_Rad_Rate_1} \\
\gamma_{\mathcal{G}\mathcal{D}} &\approx& \gamma_{\mathrm{v}} 
\,, \label{Non_Rad_Rate_2}
\end{eqnarray}
with $\gamma_{\mathrm{v}}(\omega)= 2\pi |\lambda_{ge}(\omega)|^2 \nu_{\mathrm{v}}(\omega)/\hbar^2$ the vibronic relaxation rate given by the product of the matrix-element square $|\lambda_{ge}(\omega_k)|^2$ evaluated at energy $\hbar\omega_k\equiv\hbar\omega$ and the density of states of the vibronic bath $\nu_{\mathrm{v}}(\omega)=\sum_k \delta\left( \omega - \omega_k \right)$.
From Eq.\ref{Non_Rad_Rate_2}, we see that the SSH coupling terms in Eq.\ref{H_Diss5} open a relaxation channel between the dark states manifold $\mathcal{D}$ and the ground-state $\mathcal{G}$.
Finally, the remaining Holstein-like terms in Eq.\ref{H_Diss5}, induce 
additional vibrationally-assisted relaxation rates. 
Adopting the same approximation for the diagonal matrix elements $\lambda_{e,ik}\equiv \lambda_{e} (\omega_k)$, we obtain 
\begin{eqnarray}
\gamma_{-+} &\approx& \frac{\alpha_+\alpha_-}{N} J^{(em)} \left( \Delta_{-+}\right)\gamma_\phi
\,, \label{Dephasing_Rate_1} \\
\gamma_{+-} &\approx& \frac{\alpha_+\alpha_-}{N} J^{(abs)} \left( \Delta_{-+}\right)\gamma_\phi
\,, \label{Dephasing_Rate_2} \\
\gamma_{\mathcal{D}+} &\approx& \alpha_{+} \left( 1 - \frac{1}{N}\right) 
J^{(em)} \left( \Delta_{\mathcal{D}+}\right)\gamma_\phi \,, \label{Dephasing_Rate_3} \\
\gamma_{+\mathcal{D}} &\approx& \frac{\alpha_+}{N} J^{(abs)} \left( \Delta_{\mathcal{D}+} \right)
\gamma_\phi \,, \label{Dephasing_Rate_4} \\
\gamma_{\mathcal{D}-} &\approx& \alpha_-\left( 1 - \frac{1}{N}\right) 
J^{(abs)} \left( \Delta_{-\mathcal{D}}\right)\gamma_\phi \,, \label{Dephasing_Rate_5} \\
\gamma_{-\mathcal{D}} &\approx& \frac{\alpha_-}{N} J^{(em)} \left( \Delta_{-\mathcal{D}}\right)
\gamma_\phi \,, \label{Dephasing_Rate_6}
\end{eqnarray}
with the transition energies $\Delta_{-+}=\varepsilon_+ - \varepsilon_-$, $\Delta_{\mathcal{D}+}=\varepsilon_+ - \varepsilon_{\mathrm{D}}$, 
$\Delta_{-\mathcal{D}}=\varepsilon_{\mathrm{D}} - \varepsilon_-$, and 
the dephasing rate $\gamma_\phi(\omega)= 2\pi |\lambda_{e}(\omega)|^2 \nu_{\mathrm{v}}(\omega)/\hbar^2$.
We note that the Holstein coupling terms in Eq.\ref{H_Diss5}, being diagonal in the bare (uncoupled) molecular basis, thus induce pure dephasing rates in this initial basis (contribution to the decay of off-diagonal matrix elements of the molecule density matrix). 
However, when expressed in the dressed (coupled) \textit{reacton} manybody basis, those terms become responsible for an opening of additional relaxation channels between the polariton states $\pm$ and the dark state manifold $\mathcal{D}$, as well as relaxation between upper and lower polaritons.
With respect to the nature of the initial dephasing mechanism in the uncoupled basis, we choose to  keep the convention of designing the bare rate $\gamma_\phi$ and dressed rates derived above as ``dephasing" rates.
This is in contrast to the convention used for instance in Ref.\cite{pino_quantum_2015}.
Our theoretical approach to compute the vibrational relaxation rates is consistent with 
Refs.\cite{pino_quantum_2015,Martinez_2018} which focused on the vibrational strong-coupling regime in microcavities \cite{pino_quantum_2015,Martinez_2018}.
We provide in Table \ref{table:Table3} typical values \cite{schwartz2013polariton,wang2014quantum,canaguier-durand_non-markovian_2015,bahsoun_electronic_2018} for the bare vibronic relaxation rate $\gamma_{\mathrm{v}}$, bare vibronic dephasing rate $\gamma_\phi$, as well as for the computed and dominant dressed relaxation rates obtained from Eq.\ref{Non_Rad_Rate_1} to Eq.\ref{Dephasing_Rate_6} (see blue arrows in Fig.\ref{fig:Fig6}) .
%

\section{Ultrafast reaction kinetics}
\label{Ultrafast_reaction_kinetics}
\subsubsection{Rate-equation in the \textit{reacton} basis}
\label{RE_Reacton}
\begin{figure}[tbh]
    \includegraphics[width=1.0\linewidth]{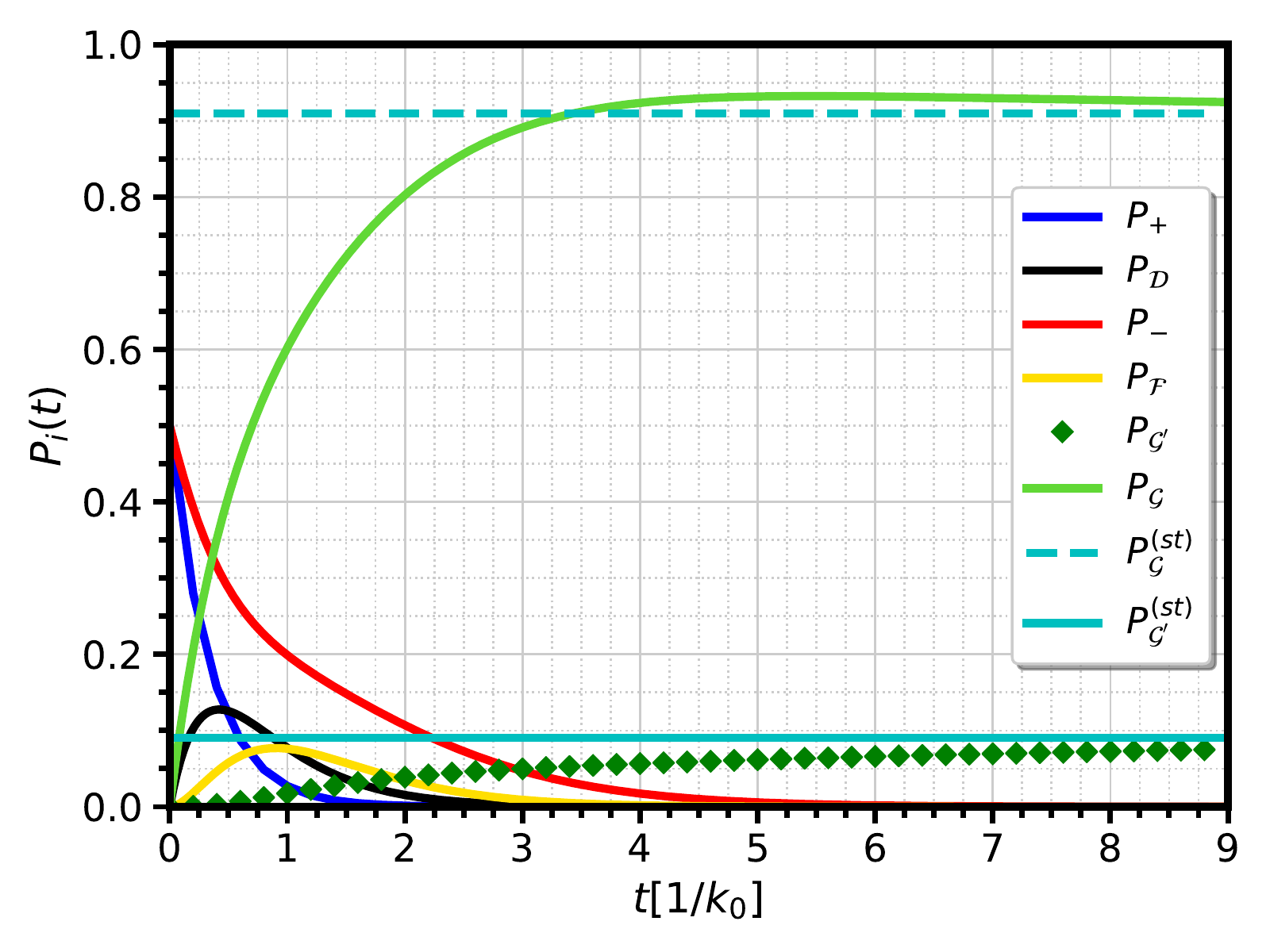}
    \caption{Probabilities $P_i(t)$ of occupying the \textit{reacton} states $i$, as a function of time $t$ in units of $1/k_{0}$ (defined in Eq.\ref{arrh}). 
    The parameters are those of Fig.\ref{fig:Fig3} case $(a)$, and Tables \ref{table:Table1}, \ref{table:Table2} and \ref{table:Table3}.
    }
    \label{fig:Fig7} 
\end{figure}
In this section, we compute the (out-of-equilibrium) occupation probabilities $P_i(t)$ as a function of time $t$ of the \textit{reacton} states $i$ involved in the whole photochemical process.
Chemical reactions (see Sec.\ref{Tuning_ET_Rate}), radiative relaxation (see Sec.\ref{Radiative_relaxation}) and non-radiative relaxation mechanisms (see Sec.\ref{Non_Radiative_relaxation})
by the environment, induce incoherent transitions amongst the
\textit{reacton} states (see arrows in Fig.\ref{fig:Fig6}).
We describe the resulting time-evolution of the populations by a rate-equation, written in the \textit{reacton} basis
\begin{eqnarray}
\dot{\underline{P}}(t) &=& \mathbb{\Gamma} \underline{P}(t)
\,, \label{RE_1} \\
\underline{P}(0) &=& \frac{1}{2}\left\lbrack 0,0,1,0,1,0 \right\rbrack
\,, \label{RE_2}
\end{eqnarray}
with $\underline{P}(t)=\left\lbrack P_{\mathcal{G}}(t), P_{\mathcal{G}'}(t), P_-(t), P_{\mathcal{D}}(t), P_+(t), P_{\mathcal{F}}(t)\right\rbrack$, the vector of populations 
$P_i(t)$, and $\mathbb{\Gamma}$ the rate-matrix with matrix-elements $\mathbb{\Gamma}_{ij}$ corresponding to the total transition rate (including chemical reaction rates, radiative and non-radiative relaxation rates) from the manybody state $j$ to the manybody state $i$.
The initial condition $\underline{P}(0)$ corresponds physically to an initial photon that has been absorbed at $t=0^-$ in order to initiate the photoreaction at $t=0^+$.
For a resonant situation $(\delta=0)$, this leads to the choice $P_-(0)=P_+(0)=1/2$ in Eq.\ref{RE_2}.
The solution of Eq.\ref{RE_1} with the initial condition of Eq.\ref{RE_2} is found by computing numerically $\underline{P}(t) = e^{\mathbb{\Gamma} t} \underline{P}(0)$.
The vector of populations can be expressed more conveniently as a linear combination of exponentially damped eigenmodes characterizing the whole photochemical process
\begin{eqnarray}
\underline{P}(t) &=&  \underline{P}^{(st)} + \sum_{\lambda \ne 0} c_\lambda \underline{v}_\lambda e^{\lambda t} 
\,, \label{RE_3} \\
c_\lambda &=& ^t\underline{w}_\lambda \: \underline{P}(0) \equiv \frac{w_{\lambda,-}+ w_{\lambda,+}}{2}
\,, \label{RE_4}
\end{eqnarray}
with $\underline{v}_\lambda$ the right-eigenvector and $^t\underline{w}_\lambda$ the left-eigenvector of the $\mathbb{\Gamma}$-matrix, associated to the real negative eigenvalue $\lambda$.
The left and right eigenvectors of $\mathbb{\Gamma}$ form a bi-orthogonal basis \cite{brody2013biorthogonal}, which enables by projection to find the unique coefficient $c_\lambda$ in Eq.\ref{RE_4} as a function of the initial condition.  
The constant vector $\underline{P}^{(st)}\equiv \underline{v}_0$ in Eq.\ref{RE_3} is the null right-eigenvector (solution of $\mathbb{\Gamma} \underline{P}^{(st)} = \underline{0}$) providing the stationary populations of the \textit{reacton} states.
We finally get for $\underline{P}^{(st)}$ and $P_{\mathcal{F}}(t)$
\begin{eqnarray}
\underline{P}^{(st)} &=& \frac{1}{k_{\mathcal{G}'\mathcal{G}}+k_{\mathcal{G}\mathcal{G}'}} \left\lbrack k_{\mathcal{G}\mathcal{G}'},k_{\mathcal{G}'\mathcal{G}},0,0,0,0 \right\rbrack
\,, \label{RE_5} \\
P_{\mathcal{F}}(t) &=&  \sum_{\lambda \ne 0}  \frac{w_{\lambda,-}+ w_{\lambda,+}}{2} v_{\lambda,\mathcal{F}} e^{\lambda t} 
\,. \label{RE_6} 
\end{eqnarray}
The stationary state in Eq.\ref{RE_5} corresponds to a chemical equilibrium between the electronic ground-state populations $P^{(st)}_{\mathcal{G}}$ and $P^{(st)}_{\mathcal{G}'}$.
%

%
\subsubsection{Time-evolution of the photoreaction}
\label{Evolution_Populations}
\begin{figure}[tbh]
    \includegraphics[width=1.0\linewidth]{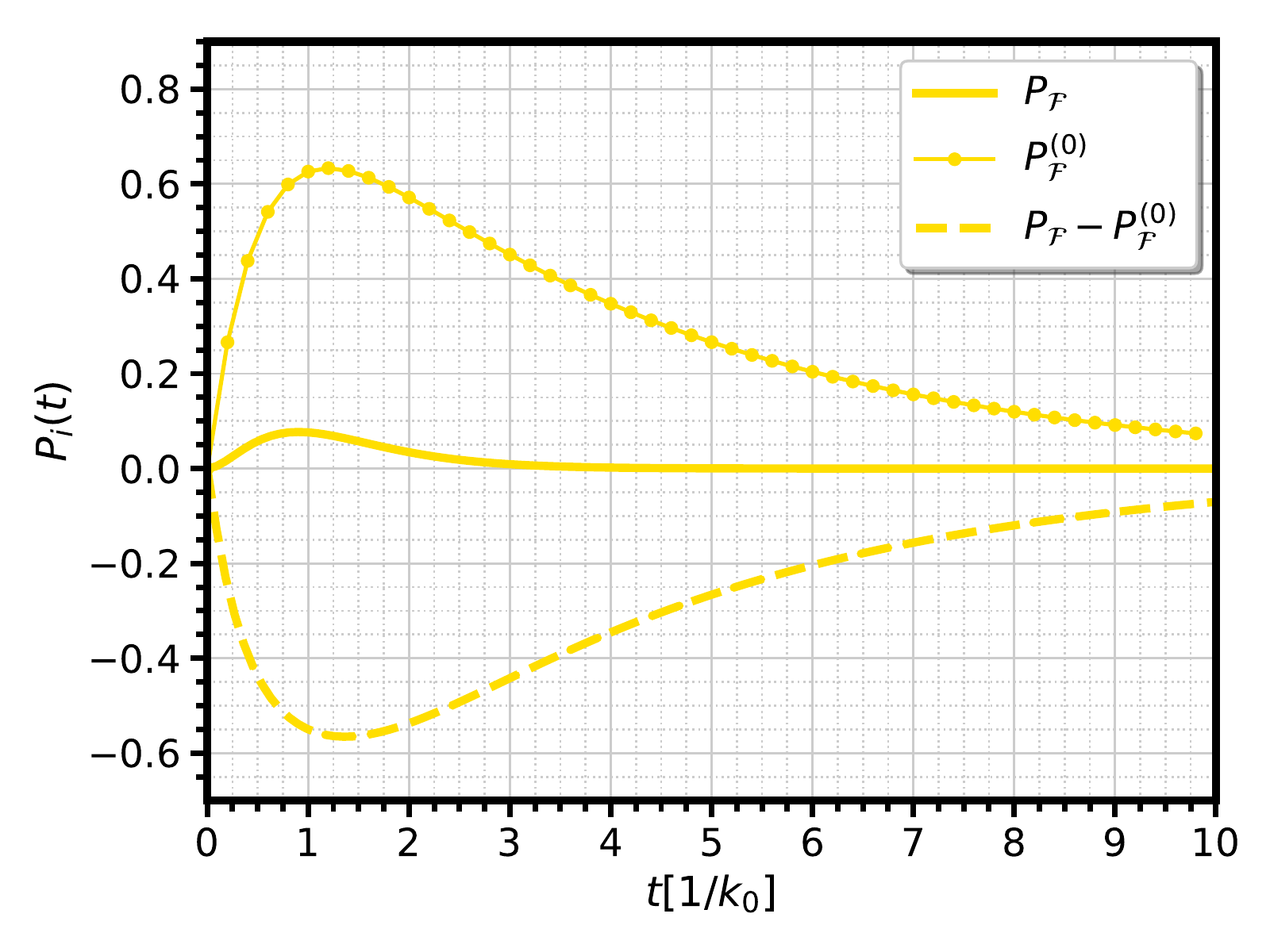}
    \caption{
    Probability $P_{\mathcal{F}}(t)$ of occupying the product-state $\mathcal{F}$ inside cavity ($\tilde{\Omega}_R=0.7\mbox{ meV}$) shown as a plain yellow curve, as a function of time $t$ in units of $1/k_{0}$ (defined in Eq.\ref{arrh}).
    The corresponding occupation probability $P^{(0)}_{\mathcal{F}}(t)$ outside cavity ($\tilde{\Omega}_R\approx 0.0\mbox{ meV}$) is shown as a dotted yellow curve.
    For comparison, we plot the difference of occupations $P_{\mathcal{F}}(t)-P_{\mathcal{F}}^{(0)}(t)$ as a dashed yellow line.
    Parameters are those of Fig.\ref{fig:Fig3} case $(a)$, and Tables \ref{table:Table1}, \ref{table:Table2} and \ref{table:Table3}.
    }
    \label{fig:Fig8} 
\end{figure}
\begin{figure}[tbh]
    \includegraphics[width=1.0\linewidth]{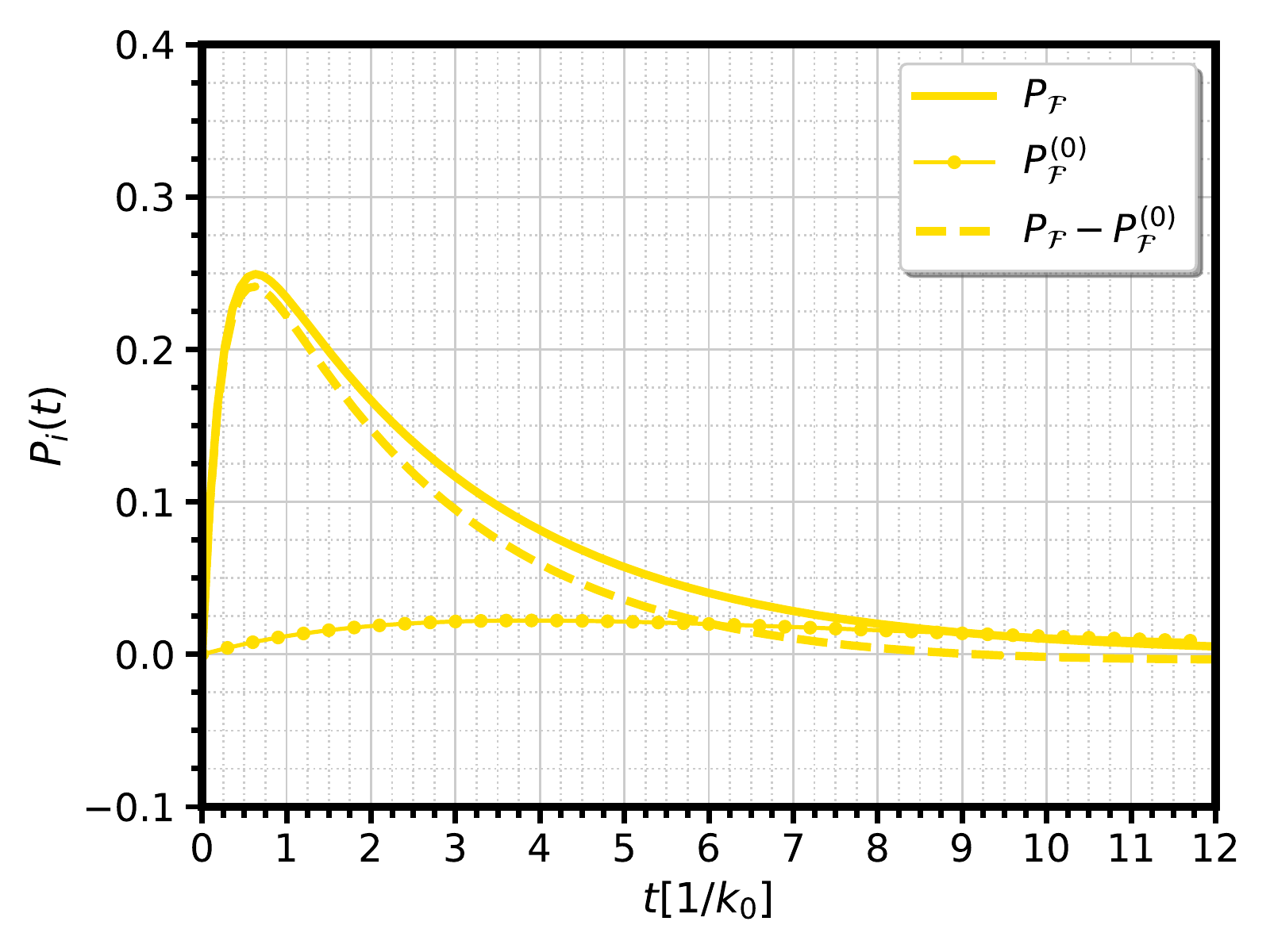}
    \caption{
    Same figure as in Fig.\ref{fig:Fig8}, but with a modified reaction driving-force $\Delta_{ef}=-0.4\mbox{ eV}$ indicated by the arrow ($b$) in Fig.\ref{fig:Fig3}.
    }
    \label{fig:Fig9} 
\end{figure}
We present in Fig.\ref{fig:Fig7} the time-evolution of $P_i(t)$, corresponding to the molecule of Fig.\ref{fig:Fig1} and case ($a$) in Fig.\ref{fig:Fig3}.
As shown in Table \ref{table:Table2} and \ref{table:Table3}, the dominating relaxation rates are the radiative ones $\Gamma_{\mathcal{G}\pm}$ (see gold downward arrows in Fig.\ref{fig:Fig6}) and the dephasing rate $\gamma_{\mathcal{D}+}$ (downward blue arrow in Fig.\ref{fig:Fig6}).
We obtain that on time scales $t \gg 1/\Gamma_{\mathcal{G}\pm},1/\gamma_{\mathcal{D}+}$, all the populations in the excited-states vanish, the stationary regime being a chemical equilibrium between the states $\mathcal{G}$ and $\mathcal{G}'$ in Eq.\ref{RE_5} (see plain green and losange green curves in Fig.\ref{fig:Fig7}).
The population of the upper polariton $P_ {+}(t)$ (plain blue curve in Fig.\ref{fig:Fig7}) is a monotonically decreasing function of time, well approximated by a single exponential decay $P_+(t) \approx e^{-\Gamma_{+} t}/2$. 
The upper polariton lifetime $1/\Gamma_{+} = 1/\left(\Gamma_{\mathcal{G}+} + \gamma_{\mathcal{G}+} + \gamma_{\mathcal{D}+}\right)$ results mainly from both optical cavity damping ($\Gamma_{\mathcal{G}+}$) and fast relaxation ($\gamma_{\mathcal{D}+}$) towards the dark-state manifold mediated by the vibrational dephasing mechanism.
The dark states thus play the role of a sink for the upper polariton state (this feature was already noticed in Ref.\cite{pino_quantum_2015}).
The population of the dark-states $P_{\mathcal{D}}(t)$ is shown as a plain dark curve in Fig.\ref{fig:Fig7}.
Its time-evolution is not monotonous but well approximated by $P_{\mathcal{D}}(t) \approx \gamma_{\mathcal{D}+} \left( e^{-\Gamma_{+} t} - e^{-\Gamma_{\mathcal{D}}t} \right)/\left\lbrack 2\left( \Gamma_\mathcal{D} -  \Gamma_{+} \right)\right\rbrack$, with the additional dark-state lifetime $1/\Gamma_{\mathcal{D}}=1/\left(\gamma_{\mathcal{G}\mathcal{D}}+k_{\mathrm{CT},\mathcal{F}\mathcal{D}}\right)$.
The existence of a maximum of $P_{\mathcal{D}}(t)$ results from a competition between the filling of the dark-state from the upper polariton with a rate $\Gamma_{+}$, and its emptying towards the ground-state $\mathcal{G}$ and excited-state $\mathcal{F}$ with rate $\Gamma_{\mathcal{D}}$.
Compared to the upper polariton, the occupation of the lower polariton $P_ {-}(t)$ (plain red curve in Fig.\ref{fig:Fig7}) is still a monotonically decreasing function of time, but with a slower rate due to the absence of ultrafast relaxation towards the dark-state manifold.
Of particular interest for photochemistry is the time-evolution of the occupation probability for the reaction product $P_ {\mathcal{F}}(t)$ (yellow plain curve in Fig.\ref{fig:Fig7}).
We show in Fig.\ref{fig:Fig8} a zoom on $P_ {\mathcal{F}}(t)$ inside the cavity ($\tilde{\Omega}_R=0.7\mbox{ meV}$ for the plain yellow curve) and the same quantity $P^{(0)}_ {\mathcal{F}}(t)$ outside cavity ($\tilde{\Omega}_R \approx 0.0\mbox{ meV}$ for the dotted yellow curve).
For our range of parameters corresponding to the reaction driving-force $\Delta_{ef}=-0.2\mbox{ eV}$ (case of the molecule in Fig.\ref{fig:Fig1} and case $(a)$ in Fig.\ref{fig:Fig3}) and choice of initial condition, we predict that $P_ {\mathcal{F}}(t) \leq P^{(0)}_ {\mathcal{F}}(t)$ at all times. 
The cavity-molecule coupling has thus an effect to slow-down the photochemical reaction compared to what is obtained outside cavity.
The same curve is plotted in Fig.\ref{fig:Fig9}, for the different value of
$\Delta_{ef}=-0.4\mbox{ eV}$ corresponding to case $(b)$ in Fig.\ref{fig:Fig3}.
In contrast to the previous case, one observes for each times that $P_ {\mathcal{F}}(t) \geq P^{(0)}_ {\mathcal{F}}(t)$, so that the effect of coupling the reactant to vacuum quantum fluctuations of the electromagnetic cavity-mode is to speed-up (and thus to enhance) the formation of the reaction product significantly, compared to the case outside cavity.
The cavity-induced slowing-down or acceleration of the appearance rate for the photoreaction product depends thus crucially on the reaction driving-force $\Delta_{ef}$ (and thus choice of the coupled-molecules), which is consistent with the analysis of the thermal CT-rate performed in Sec.\ref{Marcus_Theory_Reacton}.
The main feature observed in both Fig.\ref{fig:Fig8} and Fig.\ref{fig:Fig9}, is the non-monotonous dependence of $P_ {\mathcal{F}}(t)$ with time $t$.
We found an accurate analytical approximation of Eq.\ref{RE_6} for describing 
$P_ {\mathcal{F}}(t)$ in Fig.\ref{fig:Fig8} 
\begin{eqnarray}
P_ {\mathcal{F}}(t) &\approx& \sum_{\rho=\pm} c_{\lambda_\rho} e^{-\lambda_\rho t} 
+
c_\mathcal{D} e^{-\Gamma_\mathcal{D} t} 
+
c_\mathcal{+} e^{-\Gamma_\mathcal{+} t} 
\,, \label{Analytical_P_F_1} \\
c_\mathcal{D} &=& \sum_{\rho=\pm}\frac{\eta_\rho}{ \Gamma_\mathcal{D} - \lambda_\rho} 
\,, \label{Analytical_P_F_2} \\
c_+ &=& -\sum_{\rho=\pm}\frac{\eta_\rho}{ \Gamma_+ - \lambda_\rho} 
\,, \label{Analytical_P_F_3} \\
c_{\lambda_\rho} &=& \rho\frac{k_{\mathrm{CT},\mathcal{F}-}}{4 \mu}
+
\eta_\rho
\left\lbrack
\frac{1}{ \Gamma_\mathcal{+} - \lambda_\rho} - \frac{1}{\Gamma_\mathcal{D} - \lambda_\rho}
\right\rbrack
\,, \label{Analytical_P_F_4} 
\end{eqnarray}
with two additional decay rates $\lambda_{\rho=\pm}$ given by
\begin{eqnarray}
\lambda_\rho &=& \frac{\Gamma_{\mathcal{F}}+\Gamma_-}{2} - \rho \mu 
\,,\label{Analytical_P_F_5} \\
\mu &=& \sqrt{ \left( \frac{\Gamma_{\mathcal{F}}-\Gamma_-}{2} \right)^2 + k_{\mathrm{CT},-\mathcal{F}}k_{\mathrm{CT},\mathcal{F}-}}
\,,\label{Analytical_P_F_6} 
\end{eqnarray}
and prefactor
\begin{eqnarray}
\eta_\rho &=& 
\frac{k_{\mathrm{CT},\mathcal{F}\mathcal{D}}\gamma_{\mathcal{D}+}}{4 \left(\Gamma_\mathcal{D} - \Gamma_+\right)\mu} \left( \mu - \rho \frac{\Gamma_{\mathcal{F}}-\Gamma_-}{2} \right)
\,. \label{Analytical_P_F_7} 
\end{eqnarray}
The former expressions involve the decay-rates of the lower-polariton $\Gamma_-=\Gamma_{\mathcal{G}-}+\gamma_{\mathcal{G}-}+k_{\mathrm{CT},\mathcal{F}-}$ and
$\mathcal{F}$ excited-state $\Gamma_\mathcal{F}=\gamma_{\mathcal{G}'\mathcal{F}}+k_{\mathrm{CT},-\mathcal{F}}$. 
Compared to $P_ {+}(t)$ and $P_ {\mathcal{D}}(t)$, the time-evolution of $P_ {\mathcal{F}}(t)$ as given by Eq.\ref{Analytical_P_F_1} is more complex, as it involves 
four different relaxation time-scales ($1/\lambda_\pm$, $1/\Gamma_\mathcal{D}$ and $1/\Gamma_\mathcal{+}$).
Initially, $P_ {\mathcal{F}}(0)=0$, since the two polariton states are equally populated ($P_ {\pm}(0)=1/2$). 
At short times $t \leq 1/\Gamma_{\mathcal{G}\pm},1/\gamma_{\mathcal{D}+}$, 
the upper polaritons decays toward the dark-states manifold.
When the $\mathcal{D}$-states are significantly filled, the CT chemical reaction gets initiated, 
mainly by the dominant reaction rate $k_{\mathrm{CT},\mathcal{F}\mathcal{D}}$ (see Table \ref{table:Table1})
which is modulated by the strong light-matter coupling inside cavity.
This results in a short-time increase of the $\mathcal{F}$ product-state occupancy.
The existence of a maximum of $P_ {\mathcal{F}}(t)$ for $t\approx 1/k_0$ and a later decrease of the 
product-state occupancy, is due to the onset of the relaxation back to $\mathcal{G}'$ due to the non-radiative relaxation rate $\gamma_{\mathcal{G}'\mathcal{F}}$ (see Table \ref{table:Table3}) and to the cavity-mediated backward reaction rate $k_{\mathrm{CT},-\mathcal{F}}$ (see Table \ref{table:Table2}).
We note the importance of taking into account the losses induced by dissipation and non-radiative relaxation towards the environment in describing the photoreaction kinetics.   
The non-monotonous behavior of $P_ {\mathcal{F}}(t)-P^{(0)}_ {\mathcal{F}}(t)$ in Fig.\ref{fig:Fig8}
and Fig.\ref{fig:Fig9} is a signature of the \textit{reacton} formation, that should be observable using pump-probe spectroscopy. 
Its sign provides the information whether or not the strong-coupling of reactants to the cavity-mode enhances or inhibits the formation of the reaction product. 
There is a large room of possibilities to engineer and optimize this reaction kinetics by fine-tuning of the system parameters. 
%

\section{Conclusion and Perspectives}
\label{Perspectives}
We have investigated the chemical reactivity of solvated molecules confined inside a nanofluidic Fabry-P\'{e}rot electromagnetic cavity. 
We studied the archetypal model of a photochemical reaction for which a charge-transfer process occurs from one electronic excited-state $e$ to another excited-state $f$ of the molecule, followed by a reorganisation of the nuclei molecular conformation.
Upon tuning the cavity-frequency $\omega_c$ in resonance with the molecular transition between the electronic ground and excited states $\Delta_{ge}$, a collective polariton excitation is formed, as 
soon as the collective vacuum Rabi splitting $\tilde{\Omega}_R$ gets larger than the total losses of the cavity $\kappa$. 
We have shown that, as a result of the interaction of the molecules and cavity with the external environment, the polariton gets dressed by both intra-molecular and solvent vibrational degrees of freedom. 
We called the resulting collective excitation shared coherently between by all the reactant molecules a \textit{reacton}, by analogy with the polaron excitation in solid-state physics. 
We computed and studied in detail the modification of the polariton potential energy surfaces as well as of the equilibrium positions of the molecular vibrational modes induced by the \textit{reacton} formation. 
The former are responsible for a modification of the chemical reactivity of confined molecules compared to unconfined ones.  
We derived an extension of Marcus theory of electron-transfer reactions, taking into account the \textit{reacton} formation, and computed the kinetics of CT reaction rates for molecular populations confined in the nanofluidic electromagnetic cavity. 
We have shown the possibility to tune (acceleration or slowing down) the CT thermal reaction rate $k_{\rm{CT}}$ by changing the bare vacuum Rabi frequency $\Omega_R$, the molecule-cavity detuning $\delta$, the number of reacting molecules $N$, the driving-force of the chemical reaction $\Delta_{ef}$ and the reorganization energies $\lambda_{\mathrm{v}}$ and $\lambda_{\mathrm{S}}$. 
Our approach paves the way for new possibilities in molecular engineering, using strong-coupling of the molecules to vacuum quantum fluctuations of the electromagnetic cavity-modes. 
Finally, we derived the kinetics of the whole photochemical process, in which the CT process is one of many elementary steps.
For doing so, we had to include explicitly into the theoretical description the relaxation rates due to the optical damping of the cavity, dissipation and dephasing induced by the intra-molecular and solvent vibrational modes.
We developed for this purpose a generalized rate-equation approach expressed in the basis of manybody \textit{reacton} states, the solution of which provides the ultrafast picosecond dynamics of the 
photochemical reaction. 
Inside the cavity, we predict either an increase or a decrease of the occupation probability $P_{\mathcal{F}}(t)$ for the product-state $\mathcal{F}$ compared to outside cavity, depending on the bare reaction driving-force. 
We show that the time at which a maximum amount of reaction product is obtained, results from a delicate balance between competing environment-induced dissipation tending to decrease the net rate of product formation and the enhanced chemical reactivity due to the formation of the \textit{reacton}. 
The signature of the CT reaction should be visible in time-scales ranging from hundreds of femtoseconds to few picoseconds and in some cases to several hundreds of picoseconds \cite{patrizi2020synergistic}; these time-scales are easily attainable in regular pump-probe experiments. 
%
%

%
We assign several perspectives to extend the following paper. 
One of them is to investigate how to define properly a thermodynamical potential describing the \textit{reacton} thermodynamic properties inside the nanofluidic cavity. 
Although pioneer studies \cite{canaguier2013thermodynamics} investigated the thermodynamics of cavity-confined molecules, a proper definition and quantitative calculation of the corresponding
\textit{reacton} chemical potential is still missing. 
The former task involves to take into account into the theoretical description the spatial dependence of the cavity-mode electric field, that is responsible for spatial inhomogeneities \cite{houdre_vacuum-field_1996} in the vacuum Rabi frequency $\Omega_R$ and detuning $\delta$  experienced by each coupled molecule.
Moreover, thermal fluctuations of each molecular dipole with respect to the local electric-field direction induces the necessity to perform an additional rotational averaging \cite{craig1998molecular}, on top of the previous spatial one. 
Another interesting direction of research is to investigate the case of an open chemical reactor, namely a flow of reactants in solution that enters the optical cavity, undergoes a chemical reaction inside, and finally leaves the cavity with reaction products being collected outside.
In the case of an hydrodynamic Poiseuille flow \cite{guyon1991hydrodynamique,landau1987course}, there is a characteristic time-scale $t_{L} \approx L/4v_0$, with $L$ the longitudinal dimension of the nanofluidic cavity and $v_0 = 3 D_m/2 \rho_m$ the maximum velocity at the center of the flow ($D_m$ is the mass flow, and $\rho_m$ the liquid volumic mass).  
The ratio of $t_{L}$ to the typical time-scale of the chemical reaction $t_\chi \approx 1/k_{\rm{CT}}$, provides an adimensional parameter $\xi=k_{\mathrm{CT}} L/4v_0$.
While in our paper, the CT reaction is very fast compared to the flow velocity, thus resulting 
in $\xi \gg 1$, it would be of interest to look for other kinds of chemical reactions 
for which $\xi \approx 1$.
The former case would result in an interesting non-linear dependence of the reaction rate with the hydrodynamic flow and reactant concentration. 
We hope that our study will stimulate further theoretical and experimental investigations along those directions.
%

%
\section*{Acknowledgments}
\label{Acknowledgments}
We acknowledge financial support by Agence Nationale de la Recherche project CERCa, ANR-18-CE30-0006 and the LIGHT S\&T Graduate Program (PIA3 Investment for the Future Program, ANR-17-EURE-0021).
Initial support for this work and fruitful discussions lead in the Euskampus Transnational Common Laboratory QuantumChemPhys are acknowledged. 
%

\bibliography{references}

\end{document}